%% file: manuscript.tex
\documentclass[a4paper,fleqn,usenatbib]{mnras}
\usepackage{color}
\usepackage[T1]{fontenc}
\usepackage{ae,aecompl}
\usepackage{graphicx}
\usepackage{amsmath}	
\usepackage{amssymb}	
\usepackage{multirow}
\usepackage{microtype}

\renewcommand{\vec}[1]{\mbox{\boldmath$#1$}}

\definecolor{joerg}{rgb}{0.7, 0.4, 0.0}
\definecolor{corvin}{rgb}{0.5, 0.0, 0.5}

\usepackage{eso-pic}

\AddToShipoutPictureBG*{%
  \AtPageUpperLeft{%
    \hspace{0.75\paperwidth}%
    \raisebox{-3.5\baselineskip}{%
      \makebox[0pt][l]{\textnormal{DES 2016-0159}}
}}}%

\AddToShipoutPictureBG*{%
  \AtPageUpperLeft{%
    \hspace{0.75\paperwidth}%
    \raisebox{-4.5\baselineskip}{%
      \makebox[0pt][l]{\textnormal{FERMILAB-PUB-17-026-AE}}
}}}%
\begin{document}

\title[Weak Lensing Analysis of SPT Clusters with DES SV Data]{Weak Lensing Analysis of SPT selected Galaxy Clusters \\ using Dark Energy Survey Science Verification Data}

\input{authors}
\date{Accepted $t_2 > t_1$. Received $t_1 > t_0$; in original form $t_0$}
\pubyear{2018}

\label{firstpage}
\pagerange{\pageref{firstpage}--\pageref{lastpage}}
\maketitle

\begin{abstract}
  We present weak lensing (WL) mass constraints for a sample of massive galaxy clusters
  detected by the South Pole Telescope (SPT) via the Sunyaev-Zeldovich effect (SZE). We
  use $griz$ imaging data obtained from the Science Verification (SV) phase of the Dark
  Energy Survey (DES) to fit the WL shear signal of 33 clusters in the redshift range
  $0.25 \le z \le 0.8$ with NFW profiles and to constrain a four-parameter SPT
  mass-observable relation. To account for biases in WL masses, we introduce a WL mass
  to true mass scaling relation described by a mean bias and an intrinsic, log-normal
  scatter. We allow for correlated scatter within the WL and SZE mass-observable
  relations and use simulations to constrain priors on nuisance parameters related to
  bias and scatter from WL. We constrain the normalization of the $\zeta-M_{500}$
  relation, $A_\mathrm{SZ}=12.0_{-6.7}^{+2.6}$ when using a prior on the mass slope
  $B_\mathrm{SZ}$ from the latest SPT cluster cosmology analysis. Without this prior, we
  recover $A_\mathrm{SZ}=10.8_{-5.2}^{+2.3}$ and
  $B_\mathrm{SZ}=1.30_{-0.44}^{+0.22}$. Results in both cases imply lower cluster masses
  than measured in previous work with and without WL, although the uncertainties are
  large. The WL derived value of $B_\mathrm{SZ}$ is $\approx 20\%$ lower than the value
  preferred by the most recent SPT cluster cosmology analysis. The method demonstrated
  in this work is designed to constrain cluster masses and cosmological parameters
  simultaneously and will form the basis for subsequent studies that employ the full SPT
  cluster sample together with the DES data.
\end{abstract}


\section{Introduction}
Clusters of galaxies are the most massive collapsed objects in the
Universe. Their abundance as a function of cluster mass and redshift
is sensitive to the underlying cosmology and depends on both
the expansion history of the Universe and the process of structure
formation \citep{henry91,white93b,haiman01}. The main methods for identifying galaxy clusters include X-ray emission from the hot ($T\approx 10^8$\,K) intra-cluster medium \citep[ICM; e.g.,][]{edge90}, spatial overdensities of galaxies \citep[e.g.,][]{abell58}, and the Sunyaev-Zel'dovich effect \citep[SZE;][]{SZ}. The SZE results from the inverse Compton scattering of background Cosmic Microwave Background (CMB) photons by energetic electrons in the ICM. Although number counts of galaxy clusters constitute a powerful cosmological probe that is complementary to other
probes \citep[e.g.,][]{vikhlinin09,mantz2015,deHaan16}, there are two major obstacles for a cosmological analysis that need to be overcome.

The first obstacle is a precise understanding of the selection function. The
interpretation of number counts is limited by the knowledge of completeness
and contamination of the cluster sample to relate observed number counts to
the underlying true distribution that is predicted by cosmological
theories. The South Pole Telescope~\citep[hereafter SPT;][]{Ca11} cluster
sample has a very clean, uniform and well understood selection function that
corresponds approximately to a mass selection that is almost redshift
independent above redshifts $z\sim0.25$. The 2500~deg$^2$ SPT-SZ survey is of
sufficient depth to allow one to construct an approximately mass-limited
sample of galaxy clusters above a lower limit of
$M_{500,\mathrm{c}}\approx 3 \times
10^{14}\mathrm{M}_{\odot}$\footnote{$M_{500,\mathrm{c}}$
  denotes the mass inclosed by a sphere (radius $r_{500,\mathrm{c}}$) where
  the enclosed mean density is 500 times the critical density of the
  Universe. For convenience we also refer to these quantities as $r_{500}$ and
  $M_{500}$ in the following.} out to the highest redshifts where these
systems exist ($z\sim1.7$) \citep{bleem15}. It has been demonstrated that
cluster high frequency radio galaxies, whose emission could mask the SZE
decrement, have only a modest impact on the completeness of SZE selected
galaxy cluster samples \citep{gupta17}, and the contamination is well
described simply by noise fluctuations arising from Gaussian noise in the SPT
maps \citep{song12b,bleem15}. The SPT SZE cluster selection therefore
emphasizes the high-mass and high-redshift part of the mass function, which is
of particular interest for cosmological studies \citep[see][]{vdl10,
  Benson2013, rch13, bocquet15, deHaan16}.

The second obstacle is measuring the cluster masses. Samples of galaxy
clusters can be constructed using observables (e.g., X-ray luminosity or in
the case of SPT the significance of the SZE detection), which often also serve
as mass proxies. These mass proxies often depend on the morphological state of
the galaxy cluster and their scaling to total mass is not clear a-priori,
leading to systematic uncertainties in mass determination. To avoid biases
arising from these uncertainties, the mass-observable scaling relations need
to be calibrated against a low bias observable. Because weak lensing (WL) is
sensitive to the projected mass density, it is well suited for this task.  In
the context of SZE selected cluster samples, a number of studies have tested
the SZE based mass estimates against the WL derived masses
\citep[e.g.,][]{g1,vonderlinden2014,israel14,Hoekstra2015,ACT_WL}. These
analyses were in part motivated by an apparent tension between cosmological
constraints based on Planck CMB anisotropy and those based on galaxy clusters
\citep[][respectively]{planck2015, planck_cluster}.

To properly address the WL-calibrated SZE observable-mass scaling relation out
to intermediate redshifts with a large sample of clusters, one needs a
wide-field imaging survey of sufficient image quality over a part of the sky
imaged by an SZE survey. For this purpose, we present results from the Dark
Energy Survey~\citep[hereafter DES;][]{DES05}. DES is a large $grizY$ band
imaging survey covering a total area of 5,000 $\deg^2$ in the southern sky. It
is estimated to yield about 300 million galaxies up to $z=1.4$ when complete. The regular observations started in Fall 2013 and are planned to continue for five years. The quality and depth of the DES data are superior to any other preceding survey of similarly large footprint, in particular the Sloan Digital Sky Survey (SDSS). Prior to the main survey, a smaller area was observed to approximately full survey depth. The $\sim$200~$\deg^2$ with science quality imaging from this Science Verification (SV) period were meant as a testbed for the main survey. Because DES has by design almost complete overlap with the area observed by SPT, it is a natural choice for a weak lensing analysis of large samples of intermediate redshift SPT selected clusters where individual follow-up on larger ground or space-based telescopes would be too costly. To demonstrate the utility of DES for this task, we present a first weak lensing analysis of SPT selected galaxy clusters in the DES SV footprint.

\citet{m1} demonstrated the suitability of DES data for cluster weak lensing
using a sample of four very massive galaxy clusters and a precursor pipeline of the regular DES data processing software. A subsequent work \citep{melchior17} measured stacked shear profiles for a large sample of optically selected clusters. In this work, we will extend the WL
analysis of individual clusters to higher redshifts and lower masses
using the regular DES pipelines and data taken in regular survey mode observations. As our main goal, we will use the individual shear profiles to calibrate the mass-observable relation for SPT selected clusters of galaxies. Our method allows us to simultaneously constrain cosmological parameters and mass-observable relation parameters in a self-consistent way and can be used for larger samples of SPT selected clusters to this end.

This paper is organized as follows: in Section~\ref{sec:data} we give an overview of the DES and SPT observations as well as the associated shear catalogs and cluster sample used in this analysis. Section~\ref{sec:profiles} contains a description of the measurement of the cluster shear profiles together with the corrections we apply and tests we carry out to ensure robustness. In Section~\ref{sec:scalingrelation} we present results of our efforts to constrain the SZE observable mass scaling relation using the shear profiles from the previous section. We review our conclusions in Section~\ref{sec:conclusions}. 

Unless otherwise stated, we use a flat $\Lambda$CDM cosmology with a matter
density parameter $\Omega_\mathrm{m}=0.3089$ and a Hubble parameter $H_0 = 100\,h\,
\mathrm{km\,s^{-1}\,Mpc^{-1}}$ with $h=0.6774$, which are values extracted from a CMB analysis (TT, TE, and EE power spectra, combined with lowP and lensing) in combination with external constraints from baryon acoustic oscillations, the JLA supernova sample, and $H_0$
\citep{planck2015}. 

\section{Data}
\label{sec:data}

We provide a short overview of the entire DES program and then describe the
science verification observations and shear catalogs used in this work,
followed by a discussion of the SPT observations and the SZE selected lens
sample for this analysis.

\begin{figure}
\includegraphics[width=9cm]{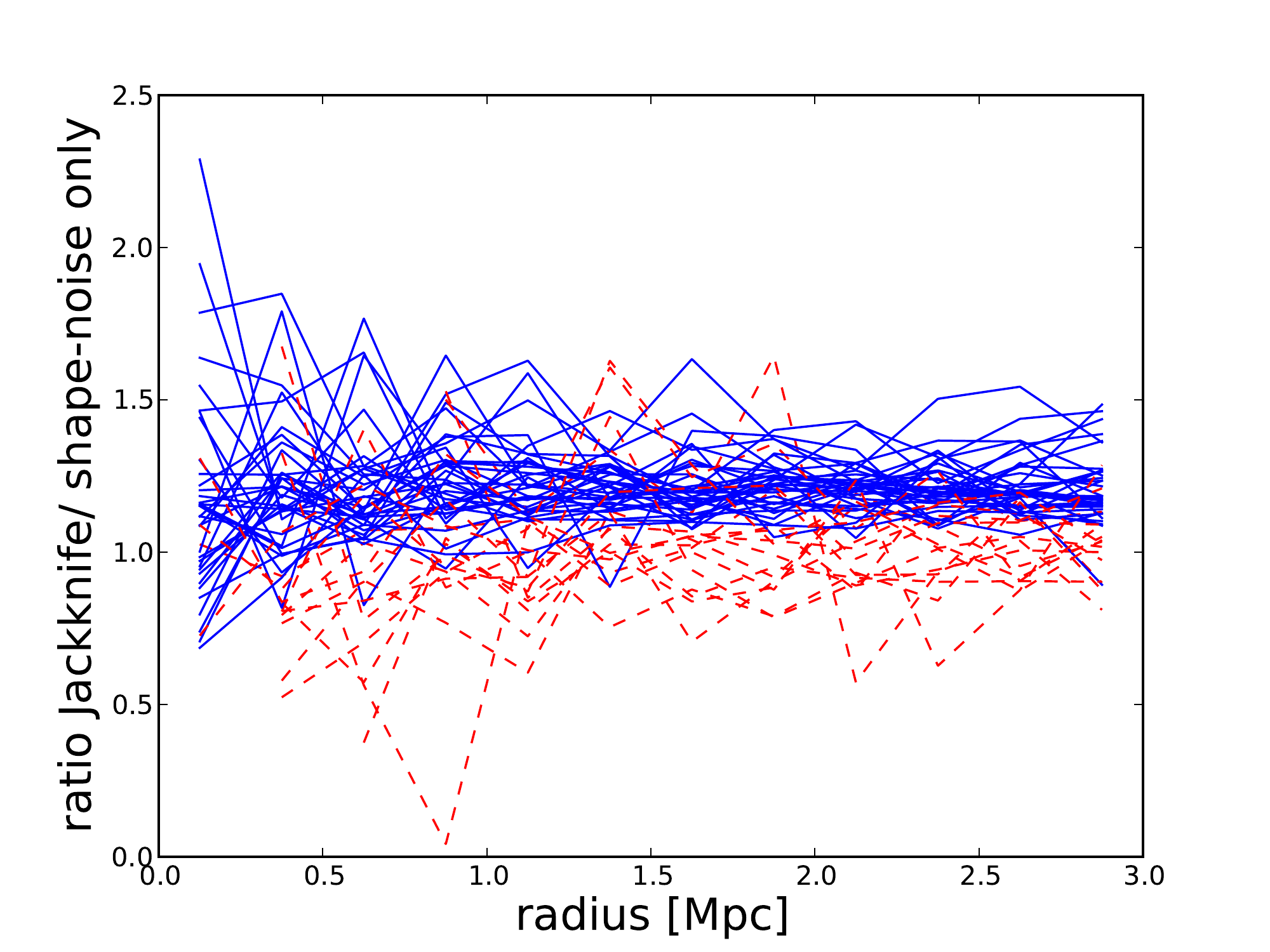}
\vskip-0.1in\caption{Ratio of Jackknife errors to intrinsic shape noise (taken to be 0.243) for the tangential shear. Each line represents an individual cluster in our sample. \textsc{ngmix} is shown in solid blue lines, \textsc{im3shape} in red dashed lines.}
\label{jkplot}
\end{figure}

\subsection{Dark Energy Survey observations}
The Dark Energy Survey \citep{DES05,DES16} is designed to yield multi-band imaging in $grizY$
bands over an angular footprint of 5,000 $\mathrm{deg}^2$. To this end, it
uses the 570 Megapixel DECam~\citep{DECam15} mounted on the 4-m Blanco
telescope at the Cerro Tololo Inter-American Observatory (CTIO). Each filter
is observed in 10 tilings of 90s exposures ($Y$-band: 45\,s during Science
Verification) over the five year survey period, and scheduling of individual
exposures employs the program \textsc{obstac}
\citep{neilsen2014}. \textsc{obstac} automatically creates the timing of
exposures based on seeing, sky brightness, and survey status.  Observations in
$riz$ bands (used for WL) are preferentially carried out in conditions of good
seeing. Additionally, deeper survey fields of 30 $\mathrm{deg}^2$ in total are
visited every 4--7 days with the main goal of measuring light curves of
supernovae. These supernova fields do not include $Y$-band imaging as part of
regular survey operations but are significantly deeper than the main survey
and visited regularly to provide finer time resolution. The survey benefits
from the very wide 3 $\mathrm{deg}^2$ field-of-view of DECam with a pixel size
of 0\farcs27. The 90\% completeness limit in each band approaches 24th
magnitude. Therefore, DES will be deeper than previous surveys of similar
solid angle like SDSS and wider than surveys of comparable depth like
CFHTLS. The median seeing is expected to be below 1\arcsec over the full
survey, and due to the addition of the $Y$-band the DES wavelength coverage
extends farther into the infrared compared to SDSS.
  

In this analysis we use Science Verification (SV) phase observations mostly obtained under regular survey conditions, and shape measurements from the $r$-, $i$- and $z$-bands, though the photo-$z$ estimates additionally rely on the $g$ band. After completion of the SV observations, the main quality cuts on the SV catalog removed the SPT-E field south of $\delta=-61^\circ$. This is the region in which the Large Magellanic Cloud resides, which has a different stellar locus than the Galaxy (affecting star-galaxy separation and photometric calibration), as well as R Doradus (the second brightest star in the infrared), which affects the photometry inside a circle of several degrees. What is more, the large number of double stars in this region complicates PSF estimation. The science-ready release of SV called `SVA1 Gold' consists of coadd catalogs that include all of these cuts and requires object detection in all four $griz$ bands. These coadd catalogs are used for object detection, flux measurements (for photo-$z$) and quality flags. 

\subsection{Dark Energy Survey shear catalogs}
\label{sec:dark-energy-survey}

The shear measurements are extracted from fitted models to all available
individual exposures for a given object after removing blacklisted exposures,
as described in \cite{im3_sv}. We use the standard SV masks
\citep{im3_sv}. These exclude circular areas around 2MASS stars and
additionally remove the 4\% of the remaining area containing a large fraction
($\approx 25\%$) of objects, whose shape could not be reliably measured. Shear measurements were performed down to
magnitude $R=24.5$ and span $139 \deg^2$ after masking in the SPT-E field.

The DES SV area is covered by shear catalogs from two shape measurement
pipelines.  We use \textsc{ngmix}\footnote{https://github/esheldon/ngmix}
\citep{ngmix}, a Gaussian mixture model fitting shear measurement code, as our
main shear measurement code. \textsc{ngmix} uses shape information from $riz$
optical bands and requires at least one valid exposure for each
band. \textsc{ngmix}, however, was not run on the entire SV footprint. For a
subsample of our lenses that is not covered by the \textsc{ngmix} analysis, we
use $r$-band catalogs from the model-fitting shear measurement code
\textsc{im3shape} instead. This is includes 5 clusters from the pointed
cluster fields.

Both codes have been shown to work well with DES SV data and produce reliable shape catalogs that pass the essential quality tests for a variety of weak lensing applications. For these and an extensive description of the DES SV shear pipeline and shape measurement codes we refer the interested reader to \cite{im3_sv}. We emphasize that the choice of \textsc{ngmix} was due to higher number densities after quality cuts, which is likely a result of using multi-band data.

The codes have been run semi-independently: though the algorithms significantly differ, they share all previous steps of data reduction, including PSF estimation and blacklisting of exposures, as outlined in \cite{im3_sv}. Both simultaneously fit to a number of single-epoch exposures for each object, instead of a fit on coadded images (where less information would be used). Galaxies have been selected according to the `Modest classifier', which uses the {\tt SExtractor} catalog parameter {\tt spread\_model} and its measurement uncertainty \citep{Bertin1996, desai12} extracted from the $i$-band image \citep[see discussion in][]{chang2015}. We remove blended objects because those are expected to have unreliable shape measurements by demanding \texttt{FLAGS\_I==0}. 

\subsubsection{Blinding}

Many scientific analyses are subject to the attempted reproduction of already published results that involves tuning the data cuts to confirm previous or expected findings \citep{blind_analysis}. We refer to this (unconscious) effect as 'observer bias'. 
 Our analysis is blinded in the following way to avoid observer bias: directly after processing and as part of the general DES shear pipeline, all shear values in the catalogs are multiplied by a hidden factor between 0.9 and 1. This acts as an effective unknown multiplicative bias that translates into an overall shift of the WL derived masses and therefore the normalization of the $M-\zeta$ scaling relation, $A_\mathrm{SZ}$. The shift due to blinding is of similar order to the mass uncertainty for the full stack, but exceeds the statistical uncertainties of cosmic shear and larger stacked lens samples that use the full SPT-E area. Only after the full analysis is fixed and all quality tests are passed, are the catalogues unblinded. However, in the process of internal collaboration review some additional tests were requested and have been carried out after unblinding.

\subsubsection{\textsc{ngmix}}
\label{sec:ngmix}
\textsc{ngmix} is a multi-purpose image fitting code. It includes a
re-implemetation of \textsc{lensfit} \citep{Miller2007, Miller2013}. In the
version used for the DES SV shape catalogs, it fits an exponential disk model
to the single-exposure galaxy images. \textsc{ngmix} fits simultaneously to
all valid exposures over the $riz$-bands and requires at least one valid
exposure in each band. It uses a shape prior from an analytical form fitted to
the ellipticity distribution of COSMOS galaxies \citep{cosmos_sn}. We use
only objects with the following quantities as reported by \textsc{ngmix}:
error flag when using the exponential model EXP\_FLAGS=0 (this includes a cut
on general \textsc{ngmix} failures, i.e. FLAGS=0), signal-to-noise
$\mathrm{SNR\_R}>10$, signal-to-noise of \textsc{ngmix} size measure
T, $\mathrm{SNR\_T\_R}>1.0$, and $0.4<\mathrm{ARATE}<0.6$. The
last item is the acceptance rate of the \textsc{ngmix} sampler and ensures
convergence of the fit. These selection parameters are relaxed from the strict
cuts suggested by \citet{im3_sv} and are based on our experience gained during
creation of the shear catalogs and expectation that due to the overall lower
source number compared to the SV cosmic shear study \citep{becker16,
  dessvcosmo16} systematic biases will remain subdominant to the increased
statistical uncertainties in this work. We will later demonstrate this
assumption to hold in Appendix~\ref{sec:shear-profile-tests}.

We use an inverse variance weight for each galaxy $i$ that takes into account
shape noise and the ($e_1$, $e_2$) covariance matrix $C$, given by
\begin{equation}
w_i = \frac {2\times\sigma_\epsilon^2} {C_{11}+C_{22}+2\times\sigma_\epsilon^2},
\label{sourceweight}
\end{equation}
where $\sigma_\epsilon=0.22$ is the shape noise contribution per component from COSMOS. We choose to use only the diagonal elements of the covariance matrix to ensure that $w_i$ is invariant under rotations.

Noise effects and choice of prior influence the observed shear and can be corrected by dividing the shear by a sensitivity that is calculated during the run of \textsc{ngmix}. Typically the shear is underestimated before this correction. Because sensitivities are noisy, we apply this correction on the ensemble of all sources used for our fitting. This is a way to estimate biases in the shape measurement algorithm in a more direct way than using external simulations. Thus, the resulting shear is effectively unbiased. This procedure is similar to the correction for noise bias in the case of \textsc{im3shape} described in the next section. 

\subsubsection{\textsc{im3shape}}
\label{sec:im3shape}
We use shape catalogs from an implementation of \textsc{im3shape}\footnote{https://bitbucket.org/joezuntz/im3shape/} \citep{im3_sv}, which was significantly improved over the version used in the simulation study of \citet{z1}. \textsc{im3shape} is a model fitting algorithm, using a \cite{deVaucouleurs1948} bulge or exponential disk model. Each object is fitted to both models, and the best-fitting model is chosen as an adequate description. The amplitude of each component is allowed to vary and may be negative, and the fitting is done simultaneously over all exposures in one band. Galaxies are selected prior to the run of \textsc{im3shape} for better performance. 

As in the case of \textsc{ngmix}, we use relaxed selection
criteria. This includes signal-to-noise \texttt{SNR > 10} and ratio of
convolved image size relative to PSF \texttt{MEAN\_RGPP\_RP>1.15}. We choose
these cuts for \textsc{im3shape} because our statistical error bars allow for
some systematic uncertainty on the overall calibration. Our choice of cuts
gives a number density (over the full SPT-E field and all redshifts) of
$n_\mathrm{g}=9.2$ $\mathrm{arcmin}^{-2}$, whereas the more conservative cuts
employed the DES-SV cosmic shear analysis \citep{becker16,dessvcosmo16} would
give $n_\mathrm{g}=5.4$ $\mathrm{arcmin}^{-2}$. We show in
Appendix~\ref{sec:shear-profile-tests} that the inclusion of these additional
galaxies leads to statistically undetectable differences in our mass calibration.

\textsc{im3shape}, as all shape measurement codes based on a maximum likelihood approach, shows systematic noise biases~\citep{noisebias}, typically expressed in terms of multiplicative bias $m_\mathrm{n}$ and additive bias $c_\mathrm{n}$ (the latter for each component separately):
\begin{equation}
e_\mathrm{obs}=(1+m_\mathrm{n})\times e_{\mathrm{true}}+c_\mathrm{n},
\end{equation}
where $e_{\mathrm{obs}}$ is the observed ellipticity and $e_{\mathrm{true}}$ is the true ellipticity of a galaxy. 
Working only with circularly averaged profiles, the additive bias is expected to average out when masking effects are negligible. The multiplicative bias however scales the tangential shear profile and therefore influences the derived masses. With help  of simulations based on galaxies from the COSMOS survey, we can express the noise bias as a function of \textsc{im3shape} signal-to-noise and galaxy size \texttt{MEAN\_RGPP\_RP}. The resulting correction is then applied to the ensemble of galaxies in a given bin.

\subsubsection{Error estimation}
If systematic effects can be neglected, the dominant source of error for a WL shear measurement comes from the intrinsic ellipticity dispersion. Therefore, in the absence of measurement noise the precision of a binned measurement of one shear component cannot be better than $\sigma_{\epsilon}/\sqrt{N_\mathrm{gal}}$ where $N_\mathrm{gal}$ is the number of source galaxies used in a given radial bin and $\sigma_{\epsilon}= 0.22$ is the intrinsic ellipticity dispersion. Because systematic uncertainties are in general hard to quantify, we use Jackknife errors as an empirical approach to estimate our measurement uncertainty on the shear profile. We calculate the signal by iteratively removing one of the used sources in each iteration. The covariance matrix for $g_+$ then can be calculated via
\begin{equation}
\mathrm{Cov}_{ij}=\frac{N-1}{N} \sum_{k=1}^N \left(g_{+,i}^k-\langle g_{+,i}^l \rangle_l\right) \left(g_{+,j}^k-\langle g_{+,j}^l \rangle_l\right)
\label{eq:cov}
\end{equation}
where $i$ and $j$ denote radial bins, and $g_+^k$ is the tangential shear without galaxy $k$. Analogous formulae are used for $g_{\times}$ and $\Delta \Sigma$ in the following. In each case we neglect off-diagonal terms for our analysis. We test the impact and determine that including the full covariance increases the mass fitted to the \textsc{im3shape} stack by about $0.1\times 10^{14}\,\mathrm{M}_\odot$ or $\approx 0.25 \sigma$, and leaves the error bars essentially unchanged.

\cite{im3_sv} calculated the shape-noise for \textsc{ngmix} in DES SV and found $\sigma_\epsilon=0.243$. Figure~\ref{jkplot} compares the Jackknife errors for the background $g_+$ with Gaussian errors assuming this value for shape-noise. Jackknife errors are larger on average by 26\% for \textsc{ngmix} and 8\% for \textsc{im3shape}, indicating that systematic errors are subdominant.

\begin{figure}
\includegraphics[width=9cm]{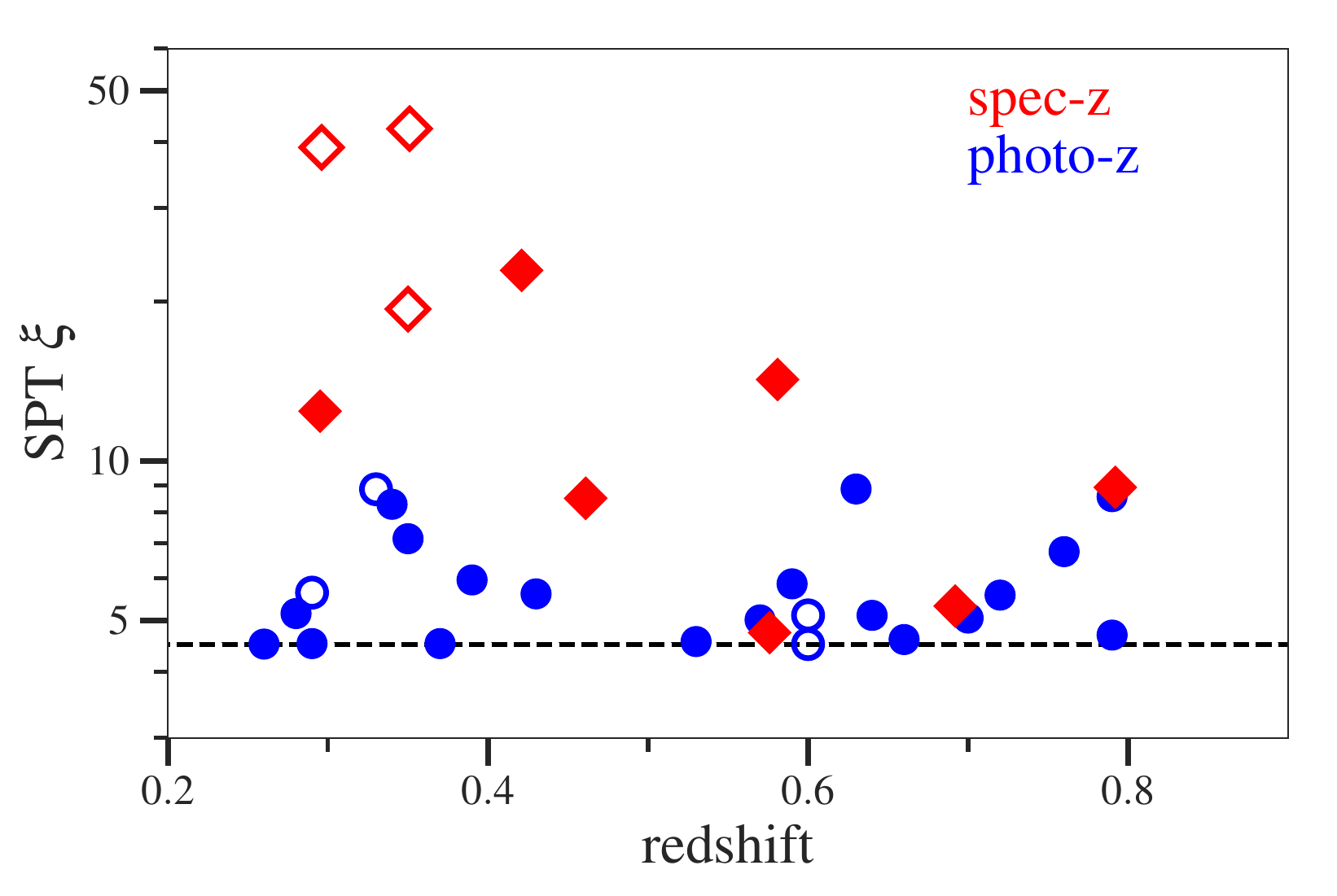}
\vskip-0.1in\caption{Our sample from the SPT-SZ catalog \citep{bleem15}. Plotted is the SPT significance $\xi$ vs. redshift. Clusters with spectroscopic redshifts are shown as red diamonds, those with only photometric redshifts as blue circles. The dashed horizontal  line corresponds to the $\xi=4.5$ limit of the catalog. Clusters covered by both shear catalogs used in this work are shown as filled symbols, those that only have shape information from \textsc{im3shape} catalogs as empty ones. As expected, most clusters lie near the catalog threshold, but the full sample spans a broad range in $\xi$.} 
\label{SPT_xi_z}
\end{figure}
Jackknife covariance matrices are often underestimated if there are too few independent samples available, and we therefore apply a correction that depends on both the number of bins and the galaxies per bin \citep{hartlap2007}. This typically increases our errors by only a few per cent.

\subsection{South Pole Telescope observations}

The SPT is a 10~m telescope located at the NSF South Pole research station. From 2007 to 2011, the telescope was configured to observe in three millimeter-wave bands (centered at 95, 150, and 220 GHz). The majority of this period was spent on a survey of a contiguous 2,500 $\mathrm{deg}^2$ area within the boundaries 20h $\leq$ R.A. $\leq$ 7h and $-65^\circ \leq \mathrm{Dec} \leq -40^\circ$. In November 2011 the observations of the whole survey area to the fiducial depth of 18 $\mu$K-arcmin in the 150 GHz band were completed. For a detailed description of the survey strategy and data processing we refer to \cite{st09}  \citep[see also][]{vdl10,wll11,moc13}. \cite{song12b} presented optical and near-infrared followup of a preliminary catalog of 720 $\mathrm{deg}^2$, including redshift estimates. The cluster catalog for the full survey area appeared in \citet{bleem15}.

Galaxy clusters are detected via their thermal SZE decrement in the 95 and 150 GHz SPT maps. These maps are created using time-ordered data processing and map-making procedures equivalent to those described in \cite{vdl10}. A multi-scale matched-filter approach is used for cluster detection \citep{mel06}, where the underlying cluster model is a $\beta$ model \citep{cavaliere76,cavaliere78} with $\beta=1$ and a core radius $\Theta_\mathrm{c}$. Twelve linearly spaced values from $0\farcm25$ to $3\farcm0$ are employed, and the observable used to quantify the cluster SZE signal is $\xi$, the detection significance maximized over this range of core radii. 

In total, 677 cluster candidates above a signal-to-noise limit of 4.5 are detected in the full SPT-SZ survey and 516 are confirmed  by optical and near-infrared imaging. This number includes 415 systems first identified with the SPT and 141 systems with spectroscopic redshift information. The median mass of this sample is $M_{500,\mathrm{c}}\approx 3.5\times 10^{14}\,\mathrm{M}_{\odot}$ and the median redshift 0.55. The highest redshift exceeds 1.4~\citep{bleem15}.

\subsection{SZE selected lens sample}
The SPT-SZ catalog has an overlap of about 100 clusters and candidates with $\xi>4.5$ over the full DES SV area, including areas that did not survive survey quality cuts in the southern part of the SPT-E field. Shear catalogs for the SPT-W field are not available at the time of this work. Some cluster candidates have not been confirmed and hence do not have a redshift estimate and are therefore excluded from this analysis. 

We restrict ourselves to clusters with redshift $0.25 < z \le 0.8$. At DES
depths, higher redshift clusters suffer from very low number densities of
lensing source galaxies and small lensing efficiency, resulting in poorly
measured, noise-dominated profiles even for the most massive systems like El
Gordo \citep{m1}. Also, complementary work with space-based HST  observations
\citep[e.g.,][]{schrabback18} is focused on providing WL based mass estimates
for systems in this redshift range. At lower redshifts the SPT selection
function is not well characterized and inclusion of clusters at $z < 0.25$
could bias our estimates of the scaling relation parameters.

This leaves us with 35 clusters with $\xi>4.5$ covered in the DES SV area. We
remove SPT-CL~J2242$-$4435 and SPT-CL~J0451$-$4952 from our lens sample
because of very low source number densities after cuts.

The remaining 33 clusters used in this analysis are listed in Table
\ref{tab:SPTclusters}, including their sky position, detection significance, core radius
$\Theta_\mathrm{c}$ and redshift. If possible we use spectrosopic redshifts (denoted by
(s)). Cluster SZE based masses $M_{500,\mathrm{SZ}}$ are taken from \citet{bleem15} and
have been derived assuming a flat $\Lambda$CDM cosmology with $\Omega_\mathrm{m}=0.3$,
$\sigma_8=0.8$ and $h=0.7$ and a fixed mass-observable relation with an intrinsic
scatter $D_\mathrm{SZ}=0.22$. These values are informational only and are not used when
deriving our scaling relation constraints.

An additional column shows the DES SV field. Most clusters are located in the SPT-E field. Several systems are in targeted cluster fields, though El Gordo is at too high a redshift to be included in our lens sample. Two systems are in one of the Supernova fields (SNE), which are deeper than the main survey. DES imaging allows optical confirmation and redshift estimates of our clusters independently of other optical follow-up observations. \cite{hennig17} identified the red sequences for SPT clusters in the SV footprint and derived comparable redshifts to those presented in \cite{bleem15} over the full redshift range. For consistency with other publications using the same SPT-SZ catalog we use the redshift estimates from \cite{bleem15} whenever possible. This is the case for almost the full sample, except for three clusters at lower signal to noise, where we employ redshift estimates and SZE based masses from \cite{saro15}.

Figure~\ref{SPT_xi_z} shows the distribution of our sample in redshift-$\xi$
space. The sample spans the full redshift range from 0.25 to 0.8, with the
majority having significance values close to the catalog threshold. Clusters
with spectroscopic redshift information are shown as red diamonds. The most
significant SPT cluster detections in our sample are in the range $0.3 < z< 0.4$, including the Bullet cluster (SPT-CL J0658$-$5556) and RXJ2248, which have been previously studied with DES data \citep{m1}. 

\input{spt_clusters_2500d}

\cite{saro15} matched SPT clusters and candidates down to $\xi=4$ to clusters identified by the optical cluster finder redMaPPer \citep{RM1} in the DES SV area, thereby confirming 5 candidates above $\xi=4.5$ and presenting redshift estimates for these systems based on their redMaPPer counterpart. We include three systems that remain after applying the SPT point source mask into our sample. \cite{bleem15} have estimated the number of false detections for $\xi<4.5$ clusters to increase from $<10\%$ at $\xi=4.5$ to $\approx 40\%$ at $\xi=4$. For the scaling relation analysis we therefore use only SPT clusters above $\xi=4.5$.

\section{Cluster shear profiles}
\label{sec:profiles}

In this section we first describe how we select the background galaxy population that is needed to construct the observed shear profiles. We then explore in Section~\ref{sec:contamination} whether the background population we have selected is contaminated by cluster galaxies.  Thereafter, we describe the theoretical profile we adopt in Section~\ref{sec:profile}, discuss the radial ranges and binning for the shear profiles in Section~\ref{sec:ranges}, and then describe the framework we introduce to account for biases and scatter in our WL mass estimates (Section~\ref{sec:sims}).

\subsection{Background source selection}
Background selection by reliable photometric redshifts has been shown to perform better than color-cuts if enough bands are available \citep[e.g.,][]{doug14}. We therefore use photometric redshifts from $griz$-bands \citep{photoz4WL} to calculate the critical surface density
\begin{equation}
\Sigma_{\mathrm{crit}} = \frac{c^2}{4 \pi G}\frac{D_\mathrm{s}}{D_\mathrm{l} D_\mathrm{ls}} \propto \frac{1}{D_\mathrm{l} \beta},
\label{sigcr}
\end{equation}
where $c$ is the speed of light, $G$ the (Newtonian) gravitational constant and $D_\mathrm{l}$, $D_\mathrm{s}$ and $D_\mathrm{ls}$ denote the angular diameter distances from the observer to the lens and the source, and from the lens to the source respectively. $\beta=D_\mathrm{ls}/D_\mathrm{s}$ is the lensing efficiency. 
 
We are using training-set based photo-$z$ estimates which have been shown to
perform better than template-based alternatives in the case of DES data
\citep{zph}. In particular, we match our shear catalogs to \textsc{skynet}
photometric redshifts~\citep{skynet2, skynet1, photoz4WL}. \textsc{skynet} is a
training set based photo-$z$ code that gives both a point estimator (the mean
or the peak of the distribution) and a full $P(z)$ distribution using
prediction trees and random forests. The training and validation sets use
28,219 and 14,317 galaxies, respectively, with measured spectra in the DES SV
footprint extending to $z=2$. Because these galaxies typically have deeper
photometry than SPT-E, they were assigned new photometric errors which were
taken from objects in the SPT-E field which are closest in a 5-d
color-magnitude space. The $P(z)$ values are tabulated for 200 values
from 0 to 1.8 and normalized to unity. The typical redshift error for
\textsc{skynet} when applied to DES SV data is $\delta z=0.08$ ($1\sigma$) for
both point estimator and $P(z)$. We choose to select our background sample by
requiring that
\begin{equation}
z_\mathrm{s} > z_\mathrm{cl}+0.2
\label{eq:zcut}
\end{equation}
holds simultaneously for both the mean and the peak of the $P(z)$
distribution. We use the former as a proxy for the source redshift
$z_\mathrm{s}$. The impact of this error for the estimation of
$\Sigma_{\mathrm{crit}}$ is described below. We construct an $N(z)$
distribution for the source sample of each cluster. If contamination by
cluster members can be neglected, $N(z)$ should not depend on cluster-centric
distance. $\beta$ is then estimated from $N(z)$ in our fitting routine for the
scaling relation. This allows us to treat the dependence of $\beta$ on
cosmological parameters in a self-consistent way. We explore the stability of
our estimation of the lensing efficiency in Section~5.4 when using a different
photometric redshift catalog.


\subsection{Cluster member contamination}
\label{sec:contamination}

Because photometric redshifts are in general noisy, cluster galaxies may scatter into the background sample. Cluster galaxies would show no shear signal from the cluster, and therefore this contamination would lead to an overall dilution of the mean shear profile and a subsequent underestimation of cluster mass. This effect can be seen as an increase in the number density of sources close to the cluster center. The radial dependence of the number density profile is also affected by magnification and the obscuration of the sky by bright foreground objects. Masking of, e.g., bright stars (including the 2MASS catalog), image artifacts or because of survey edges also must be taken into account to derive correct number densities. Noting that magnification only contributes significantly in the very inner regions \citep{chiu16b}, which we neglect in our shear analysis, we leave this effect uncorrected (but see \citet{schrabback18} for an investigation of its potentially larger impact for clusters at higher redshift).

\subsubsection{Radial trend in background density}
To estimate a correction for the contamination we first assume that the contamination by cluster galaxies decreases with increasing distance $r/r_{500,\mathrm{SZ}}$ from the cluster center, where the scale radius is set by the cluster mass (as given in \citet{bleem15}). Following \citet{doug14}, we model the effects of the contamination on the background number density as
\begin{equation}
n_\mathrm{corr}( r)= n_0 \times \left( 1+ f_{500}\exp{[1-r/r_{500,\mathrm{SZ}}]} \right) 
\end{equation}
where $n_0$ denotes the uncontaminated background number density which is a constant and $f_{500}$ is the contamination fraction at a cluster-centric distance $r_{500,\mathrm{SZ}}$. We perform a simultaneous fit for a global $f_{500}$ and a different $n_0$ for each cluster.

\begin{table}
\centering
\caption{Cluster member contamination constraints (evaluated at $r_{500,\mathrm{SZ}}$) extracted from the various subsamples.}
\begin{tabular}{lcc}
\hline\hline
& \textsc{ngmix} & \textsc{im3shape} \\
Subsample & [\%] & [\%] \\
\hline\hline
full bg   & $8.1 \pm 6.9$  & $9.3 \pm 7.0$ \\
low $z_s$  & $12.1 \pm 6.9$ & $9.5 \pm 7.6$ \\
mid $z_s$  & $2.6 \pm 7.6$  & $1.9 \pm 6.6$ \\
high $z_s$ & $0.9 \pm 7.9$  & $2.7\pm 8.4$  \\
low $z_l$  & $8.1 \pm 6.9$  & $10.7 \pm 8.9$ \\
high $z_l$ & $4.6 \pm 13.5$ & $1.4 \pm 15.0$ \\
\hline\hline
 \end{tabular}
\label{tab:Contam}
\end{table}

\begin{figure}
\includegraphics[width=\columnwidth]{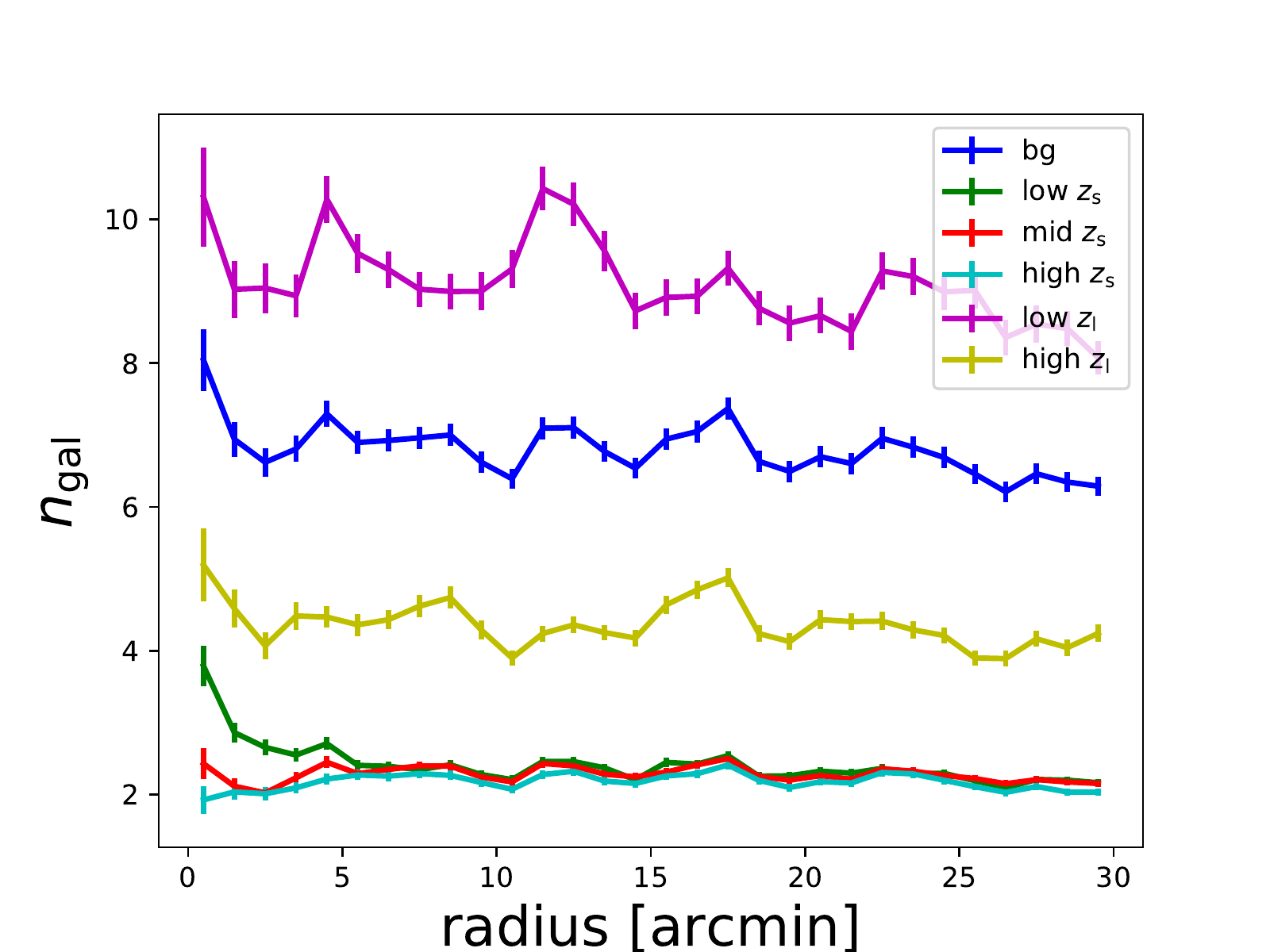}
\vskip-0.1in\caption{Number density profile of our source galaxy sample from \textsc{ngmix} as a
  function of cluster-centric angular distance.  The full sample is shown in blue, and
  three different slices in increasing source redshift are visible in green, red and
  cyan. The full source samples for low-$z$ and high-$z$ lenses are shown in magenta and
  yellow. This figure is for illustration only, because the contamination is evaluated
  for individual clusters rather than the stack shown above. The error bars are the
  Poisson errors of the number counts.}
\label{fig:contam}
\end{figure}

Figure~\ref{fig:contam} shows the average number density profile of our
\textsc{ngmix} sources as a function of cluster centric distance, including
splits in source and lens redshift. Table~\ref{tab:Contam} summarizes our
estimates of contamination.  We find a value of
$f_{500}=(8.1 \pm 6.9)$~percent for the full sample of \textsc{ngmix} sources
and lenses (in blue), very close to no contamination, and
$(9.3 \pm 7.0)$~percent for the \textsc{im3shape} sources. Without redshift
selection (equation~\ref{eq:zcut}) we would get $(11.3 \pm 2.1)$~percent.

Splitting the sources for each cluster into three equally populated source
redshift bins (green, red, cyan) shows a lot of fluctation but no significant
contamination for any bin. Splitting the cluster sample at the median lens
redshift also gives values of $f_{500}$ consistent with zero (magenta and
yellow lines) at the $1.2\sigma$ level.
 
Additionally, a small $f_{500}$ would not affect our conclusions, given the
large statistical uncertainties in our current analysis. Therefore, we choose
not to correct the tangential shear signal. Indeed, no significant cluster
contamination is expected, because we use photometric redshifts and a
background selection that corresponds to $\approx 2.5 \times \delta z$ above
the cluster redshift.

\subsubsection{P($z$) decomposition}
\label{sec:P_of_z_decomposition}
As a cross-check for our contamination correction we use an adaptation of the method described in \cite{g1} in the case of individual source redshift distributions. Because this method does not use number densities from our source catalog, it is subject to different systematics.

\begin{figure}
\vskip-0.1in
\includegraphics[width=8.7cm]{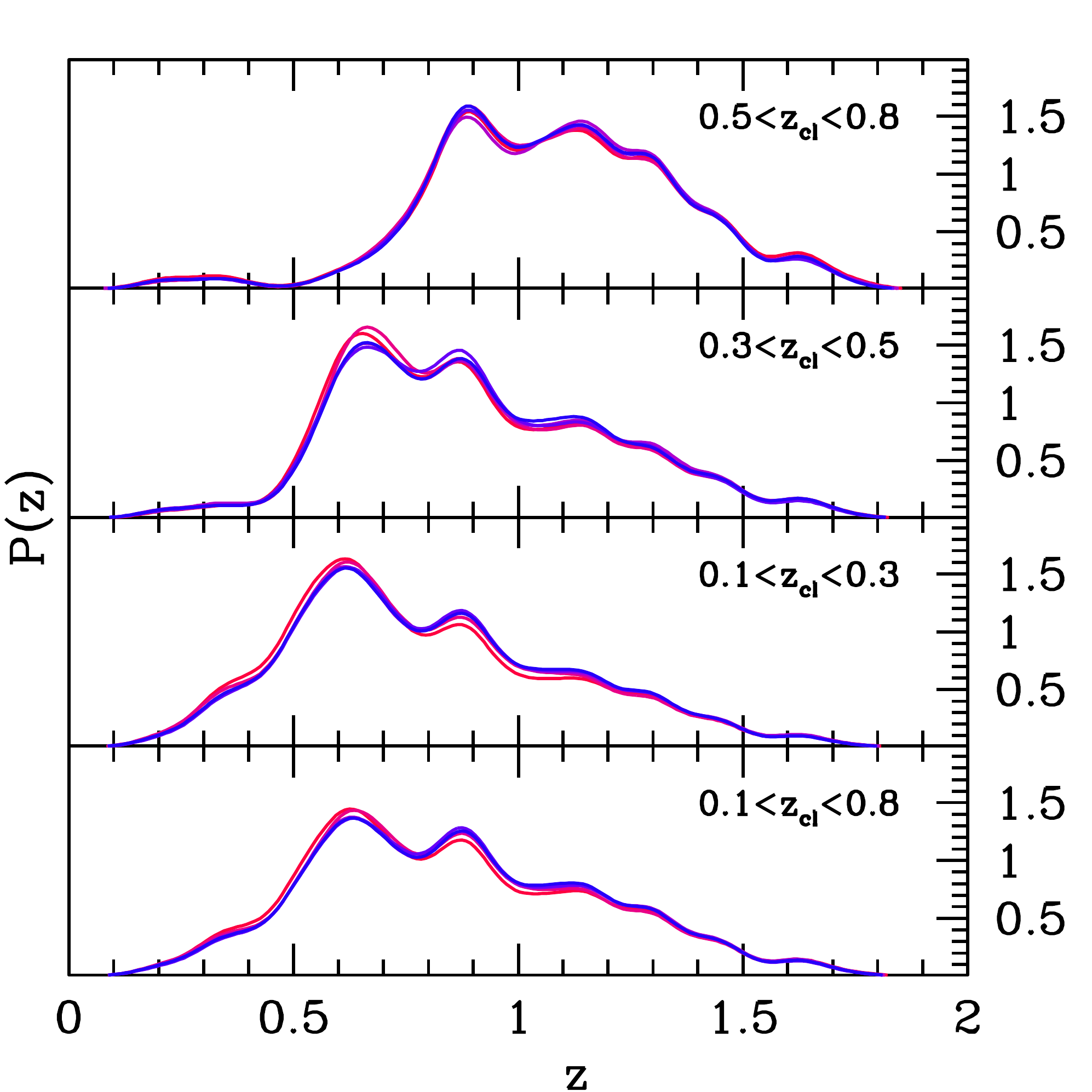}
\vskip-0.10in
\caption{P($z$) distribution of \textsc{ngmix} sources. We split the source population
  into 5 radial bins, ranging from red (innermost) to blue (outermost). The bottom panel
  shows the stack of all clusters, and the 3 top panels show slices in lens redshift. We
  estimate an overall contamination of ($3\pm1$)~per cent in the two inner bins (see
  discussion in section~\ref{sec:P_of_z_decomposition}).}
\label{fig:daniel}
\end{figure}

We summarize this method briefly and refer the interested reader to the original paper \citep{g1} and a study of the stacked WL signal from redMaPPer clusters in DES SV data \citep{melchior17} for its adaptation to DES $P(z)$'s. The source galaxy redshift distribution is modelled with two components: a spatially constant background and a radially varying contaminant of cluster galaxies. Comparing the $P(z)$'s in radial bins around the cluster center with a local background at large separation allows one to infer the level of contamination needed to recover the observed radial change in the $P(z)$ distribution. 
We choose five equally populated radial bins from 0.75--2.5~Mpc and find an overall contamination of $3 \pm 1 \%$ in the two innermost bins, translating to $f_{500}=(3.8 \pm 1.3)$~percent\footnote{We note that a direct decomposition was not possible because the $P(z)$ distribution depends only very weakly on the radius. Instead we looked at differences in the cumulative redshift distribution between radial bins.}. Fig.~\ref{fig:daniel} shows the radial dependence of the $P(z)$ distribution for the full source sample and three slices in lens redshift.

Although both methods give consistent results for the scale of the contamination, the $P(z)$ decomposition approach provides higher significance due its smaller measurement errors. We find in a similar analysis \citep{dietrich19} that this level of $f_{500}$ translates to a $\approx 2$~percent shift in mass, which is about an order of magnitude smaller than our statistical error.

\subsection{Assumed cluster profile}
\label{sec:profile}

Simulations have shown that the profile of a dark matter halo is on average well
approximated by an Navarro-Frenk-White (NFW) profile~\citep{NFW}
\begin{equation}
\rho_{\mathrm{NFW}}=\frac{\rho_0}{(r/r_\mathrm{s})(1+r/r_\mathrm{s})^2}
\end{equation}
which has two free parameters $\rho_0$ and $r_\mathrm{s}$, although more recent work
indicate that the \citet{einasto65} profile is a better fit for massive clusters
\citep[][and references therein]{klypin16}. We will calibrate the impact of deviations
from a spherical NFW profile using simulations (cf. Section~\ref{sec:sims}).

Because we are interested in the mass $M_{\Delta,\mathrm{c}}$ residing within a sphere
of radius $r_\Delta$ with an average overdensity that is $\Delta$ times the critical
density of the Universe at the cluster redshift, it is convenient to rewrite the
NFW-profile using $M_{\Delta,\mathrm{c}}$ and concentration
$c_{\Delta,\mathrm{c}}=r_{\Delta,\mathrm{c}}/r_\mathrm{s}$ as a parametrization.  For
the scaling relation analysis we use $\Delta=500$, because this will simplify comparison
to previous results.

An analytic expression for the radial dependence of the tangential shear for an NFW density profile has been presented elsewhere \citep{Bartelmann1996,nfwlens}. We use this result in our weak-lensing analysis. Because our WL data barely constrain the concentration, we adopt a concentration from previously published mass-concentration relations extracted from simulations \citep{DK15}. We find by comparing to another relation \citep{duffy08} that our results do not depend on this choice (see Section~\ref{sec:results}).

\subsection{Radial fitting range and binning}
\label{sec:ranges}

Masses derived from a WL analysis may show percent level biases depending on both the inner and outer radius of the fit region \citep{beckerkravtsov}. Excluding the central region suppresses the influence of miscentering, concentration, baryonic effects on the halo profile and a departure from the pure WL regime. On the observational side, deblending, neighbour effects and contamination by cluster galaxies degrade the reported shears for small cluster-centric distances. At large cluster-centric distances the signal is dominated by the 2-halo term and potentially by uncorrelated structures along the line of sight and the profile is not well-described by an NFW profile. To minimise the impact of these biases, we fit in the radial range from 750 kpc--2.5 Mpc for our reference cosmology, which corresponds roughly to $0.5\text{--}2\, r_{500}$ for a halo of mass $M_{500} = 3 \times 10^{14}\,\mathrm{M}_\odot$.

Because the number of sources after our cuts differs significantly from cluster to
cluster (due to depth variations after cleaning and the large span in lens redshifts),
we adopt an adaptive binning scheme where we have at least 5 bins but for background
samples larger than 1,000 galaxies, we divide the sample by 200 and take the truncated
result to be the number of bins. We tested a variety of binning schemes and found that
the choice of binning employed does not systematically influence our results.

The input data to our analysis are (i) the cosmology independent tangential shear profiles, (ii) the associated uncertainties as described in equation~(\ref{eq:cov}) and (iii) the source redshift distributions $N(z)$ weighted by the shear weight of our source sample. Whenever possible we use the \textsc{ngmix} shear catalog, because it has higher number-densities and larger numbers of exposures per object. For nine clusters, mainly outside of SPT-E, we rely on the \textsc{im3shape} shear catalog. Table~\ref{tab:SPTWL} shows the number of galaxies used for our fit and the derived number of bins  for both catalogs.

\input{spt_clusters_2500d_wl}

\subsection{Calibration of WL mass bias and scatter}
\label{sec:sims}
In our analysis we use the cluster center derived during the SZE detection process as the shear profile center. The SZE center scatters about the BCG location \citep{song12b} in a manner consistent with the scatter of the X-ray center about the BCG location \citep{lin04b}, once the additional positional uncertainties from the SPT beam are taken into account. Similar results are found in the scatter of the SZE position around the cluster optical centers \citep{Saro2014}. Studies of simulated cluster ensembles show that the offset distribution between the true center of the cluster potential and the SZE center behave similarly to these observations involving the BCG positions \citep{gupta17}. Measuring shear profiles around a position that is offset from the true center of the cluster potential will tend to decrease the shear signal at small radii and hence result in an underestimate of the weak lensing mass.  This effect has to be accounted for to obtain accurate cluster masses.

In addition, other effects such as our choice of the projected NFW model and the radial range we use to carry out the fitting also impact the accuracy and precision with which we can estimate the underlying halo mass from the weak lensing mass.  In addition,  large scale structure surrounding the cluster could potentially lead to biases in our WL masses, and the unassociated large scale structure along the line of sight toward the cluster could introduce additional scatter in our measurements.
 
To allow for the fact that the WL masses $M_\mathrm{WL}$ we measure would in general be biased and noisy probes of the underlying true cluster mass $M_\mathrm{true}$
within $r_{500,\textrm{c}}$ that we seek to measure, we introduce a simple linear relationship the weak lensing and tru masses
\begin{equation}
M_\mathrm{WL} = b_\mathrm{WL}  M_\mathrm{true},
\label{eq:wlbias}
\end{equation}
where $b_\mathrm{WL}$ is a bias parameter.  In addition, we add a scatter parameter
$\sigma_\mathrm{WL}$, which quantified the intrinsic scatter of the weak lensing mass at fixed true mass.
With these two additional degrees of freedom we can then include estimates for the characteristic bias and scatter of our weak lensing masses.  As described in details in Appendix~\ref{sec:wlpriors}, we then use mock observations of simulated galaxy clusters to 
understand the bias and scatter in the WL mass.  Results of this study lead to priors on these two
parameters, as justified in Appendix~\ref{sec:wlpriors}, that are $b_\mathrm{WL}=0.934\pm0.04$ and $\sigma_\mathrm{WL}=0.25\pm0.12$.  The uncertainty on the mean bias could be further reduced through studies of larger samples of mock observations, but the level of this ``theoretical'' uncertainty on the bias is already much smaller than the uncertainties associated with the shear multiplicative bias, the photometric redshift bias and the cluster contamination.  These biases are listed separately in Table~\ref{tab:sysl} and sum in quadrature to a total uncertainty of 0.18 that is adopted for the uncertainty on the weak lensing bias parameter in Table~\ref{tab:ScalRel}.

\input{sys_table}

\section{Scaling Relation Analysis}
\label{sec:scalingrelation}
In this section, we describe the analysis method to derive the scaling relation parameters and present the results.  We present the Bayesian framework in Section~\ref{sec:bayesian}, detail the priors in Section~\ref{sec:priors} and then present our results with a comparison to prior work in Sections~\ref{sec:results} and~\ref{sec:discussion}.

\subsection{Bayesian foreward modeling framework}
\label{sec:bayesian}

The freedom to maximize the SPT significance $\xi$ across three parameters (right
ascension, declination, and core radius $\Theta_\mathrm{c}$) in the presence of a noise
field will tend to raise the amplitude of the observed peak. That is, the ensemble
average of $\xi$ across many noise realizations, $\langle\xi\rangle$, will be boosted by
some amount as compared to an unbiased significance $\zeta$, which is measured without
these degrees freedom \citep{vdl10}. It can be
estimated for $\zeta> 2$ by
\begin{equation}
\zeta = \sqrt{\langle \xi \rangle^2-3}\;.
\label{eq:xizeta}
\end{equation}

The unbiased significance $\zeta$ can be related to the mass enclosed by a 
sphere with a mean overdensity of
500 times the critical density of the Universe, $M_{500,\mathrm{c}}$, by the
mass-observable relation
\begin{equation}
\zeta = A_\mathrm{SZ} \left( \frac{M_{500,\mathrm{c}}}{3 \times 10^{14}
    \mathrm{M}_{\odot} h^{-1}}\right) ^{B_\mathrm{SZ}}
\left(\frac{E(z)}{E(0.6)}\right)^{C_\mathrm{SZ}}, 
\label{eq:SZmor}
\end{equation}
where $A_\mathrm{SZ}$ is the normalization, $B_\mathrm{SZ}$ the mass slope,
$C_\mathrm{SZ}$ the redshift evolution, and $E(z) = H(z)/H_0$ \citep{vdl10}. An
additional parameter $D_\mathrm{SZ}$ describes the intrinsic scatter in $\zeta$ which is
assumed to be log-normal and constant as a function of mass and redshift. 

Power law scaling relations among cluster observables that exhibit low intrinsic scatter were 
first discovered in the X-ray \citep{mohr97} and immediately interpreted as evidence that
observable properties of clusters scale with the underlying cluster halo mass.  These scaling relations (observable to observable, observable to mass)
were apparent in clusters from hydrodynamical simulations of the time 
but with the wrong mass trends.  It was
quickly apparent that the mass trend of ICM based observables depends on the thermodynamic history of the ICM, which is impacted by feedback from star formation and AGN.  The existence of these early
X-ray scaling relations \citep[see also][]{mohr99} already implied the existence of SZE scaling relations of similar form, although direct observation at that time was not possible.

The first observations of the SZE scaling relations were enabled through the SPT sample and with input from followup X-ray observations with Chandra \citep{andersson11}.  Detailed analysis of the expected distribution of scatter and the redshift evolution of the SZE scaling relations have been studied with simulations \citep[see, e.g.,][]{gupta17}.  Finally, within the last three cosmological analyses of the SPT selected sample, the above scaling relation has been adopted and goodness of fit tests have been carried out.  To date, starting first with a sample of 100 clusters and moving now to a sample of 400 clusters, there have been no indications of tension between the observations and this underlying scaling relation \citep{bocquet15,deHaan16,bocquet18}.  Future, larger sample will of course allow more stringest tests and will likely lead to the need for additional freedom in the functional form \citep[see, e.g., a similar analysis framework set up for the much larger eROSITA cluster sample that does indeed require additional parameters;][]{grandis18}.

The parameter values of $B_\mathrm{SZ}$ and $C_\mathrm{SZ}$ in the relation above are therefore impacted by the thermodynamic history of the ICM and cannot be predicted with precision even with the latest generations of hydrodynamical simulations.  Importantly, to obtain unbiased cosmological results, one must introduce these degrees of freedom in the astrophysical scaling relation and then constrain them with the use of weak lensing masses.  This is indeed the goal of our analysis.  For reference, self-similar scaling in mass and redshift for the cluster population would correspond to values of approximately 
$B_\mathrm{SZ} \approx 1.3$ and $C_\mathrm{SZ} \approx 0.7$.

To constrain the four parameters in this model, both simulation priors and X-ray and
velocity dispersion information for a subset of the SPT clusters have been used. Recent
calibration studies \citep{bocquet15, deHaan16} simultaneously fitted cosmological
parameters to take into account the cosmological dependence of the scaling relation and
the observational mass constraints.

To constrain the $\zeta\text{--}M$ scaling relation given above, we use an extension of
the analysis code developed in \citet{bocquet15}. The observational constraints include
(i) the tangential shear profiles for individual clusters and (ii) the redshift
distribution $N(z)$ of source galaxies. We choose these two quantities instead of a
combined $\Delta \Sigma$ profile, because the latter is cosmology dependent, and we want
to isolate all cosmological dependencies when pursuing either a cosmological or mass
calibration analysis \citep[e.g.][]{majumdar03,benson13,bocquet15}.  An example of the shear profile 
(but in this case stacked over the whole sample) appears in Fig.~\ref{fig:im3comp}.

We use a Bayesian framework to estimate the likelihood of each cluster in our sample
forward modelling from observed cluster detection significance $\xi$ to the probability
of finding the observed shear profile
\begin{equation}
  \begin{split}
    &P\left(g_{+,i}|N_i(z),\xi_i, z_i, \vec{p} \right)= \\
    &\int \mathrm{d}M_\mathrm{WL} P\left(g_{+,i}|N_i(z), M_\mathrm{WL}, z_i,
      \vec{p} \right) P \left(M_\mathrm{WL}| \xi_i, z_i,
      \vec{p}\right),
  \end{split}
  \label{eq:13}
\end{equation}
where $i$ runs over all clusters in our sample. 
To be more explicit, the model first computes how probable a given
cluster weak lensing mass is, given the observables $\xi$ and $z$ and the model parameters $\vec{p}$,
which include the scaling relation parameters. This is the last factor on the right hand
side of eq.~(\ref{eq:13}). We then compute the probability of measuring the tangential
shear we have in our data for a given cluster weak lensing mass, cluster redshift, shear galaxy redshift
distribution $N(z)$, and model parameters $\vec{p}$. This is the
first factor in the integral in eq.~(\ref{eq:13}), and it is computed for each radial bin
assuming that the errors on the tangential shear are the standard deviation of a
Gaussian distribution. To obtain a single scalar probability distribution
$P\left(g_{+,i}|N_i(z), M_\mathrm{WL}, z_i, \vec{p} \right)$, the probabilities of all
radial bins are multiplied. 

This forward modeling approach has the benefit that we can naturally deal with clusters
whose measured shear profiles would be consistent with zero weak lensing mass or 
that exhibit negative shear in some radial bins. It is also flexible enough to deal with cluster
catalogs in which only a subset of the clusters have follow-up data, such as
weak lensing observations, in a self-consistent and unbiased way as long as the selection of follow-up observations is not correlated with SZE properties.

We correct for Eddington bias by weighting by the mass function $P(M_\mathrm{true}|z)$
\citep{tinker2008,bocquet16} when calculating $P \left(M| \xi_i, z_i,
  \vec{p}\right)$. This is necessary, because we select clusters by requiring that their
SZE detection significance satisfies $\xi > 4.5$, which directly relates to $\zeta$ in
our scaling relation via equation~(\ref{eq:xizeta}). 
The logarithm of our full likelihood is then given by
\begin{equation}
\ln \mathcal{L} = \sum_{i=1}^{N_\mathrm{cluster}} \ln P\left(g_{+,i}|\xi_i,
  z_i, \vec{p} \right) + const.
\end{equation}
Using the mass function in this way lets us calibrate the mass-observable relation
without self-calibration using the information present in cluster number counts. On
the one hand, we can straightforwardly extend this likelihood function to include other
observables, e.g. X-ray data \citep{dietrich19}, and scaling relations. On the other
hand we can include an additional term incorporating cluster abundance (Bocquet et al.,
in prep.). Future cosmological analysis using the framework presented here will make use
of cluster number counts to constrain cosmological parameters, self-calibrate
mass-observable relations, and concurrently calibrate the normalization and evolution of
these scalings.

As already mentioned, we allow for departures between the WL and true masses from either
systematics or intrinsic scatter using equation~(\ref{eq:wlbias}) with an (intrinsic)
scatter $\sigma_\mathrm{WL}$. Additionally, $\sigma_\mathrm{WL}$ and $D_\mathrm{SZ}$ may
be correlated, and so we include a correlation coefficient $\rho_\mathrm{SZ-WL}$.  In
our analysis we are then simultaneously fitting the following 7 parameters:
$\vec{p}=\{ A_\mathrm{SZ}$, $B_\mathrm{SZ}$, $C_\mathrm{SZ}$, $D_\mathrm{SZ}$,
$b_\mathrm{WL}$, $\sigma_\mathrm{WL}$, $\rho_\mathrm{SZ-WL} \}$.

We discard a burn-in phase that corresponds to five times the auto-correlation length
and consider our chains converged if the \citet{gelman1992} convergence diagnostic
$\hat{R}<1.1$.

\subsection{Priors}
\label{sec:priors}

In contrast to previous analyses \citep[e.g.][]{bocquet15, deHaan16}, we adopt
a flat prior on $\ln A_\mathrm{SZ}$ instead of a flat prior on
$A_\mathrm{SZ}$. This is motivated by the linear form of our SZE observable
mass relation (equation~(\ref{eq:SZmor})) in log-space. The uninformative
prior on the intercept of a line is flat and transforming this back to the
power law relation (\ref{eq:SZmor}) leads to a prior that is proportional to
$1/A_\mathrm{SZ}$. Inded our experience confirms that in the limit of lower
number densities, i.e. lower SNR, the prior becomes more dominant and a flat
prior on $A_\mathrm{SZ}$ biases the results towards high values. This bias
is removed by our choice of prior. Similarly, we adopt an uninformative prior
on $B_\mathrm{SZ}$ proportional to $(1+B_\mathrm{SZ}^2)^{-1.5}$. This
corresponds to a flat prior on the angle of the line rather than its slope. We
refer the interested reader to the original publication \citep{jaynes1983} for
a more detailed discussion of this choice.  We use the following Gaussian
priors on the other scaling relation parameters:
$B_\mathrm{SZ}=1.668 \pm 0.083$, $C_\mathrm{SZ}=0.550 \pm 0.315$ and
$D_\mathrm{SZ}=0.199 \pm 0.069$, which correspond to the
$\mathrm{SPT}_\mathrm{CL}$ constraints presented in the latest SPT cluster
cosmology analysis \citep{deHaan16}. These constraints adopted external priors
on $H_0$ and on $\Omega_\textrm{b}$ from Big Bang
Nucleosynthesis. Additionally, we assume a flat prior on
$\rho_\mathrm{SZ-WL} \in [-1,1]$. These values are listed in
Table~\ref{tab:ScalRel}. We also use the prior for $b_\mathrm{WL}$ derived in
Section~\ref{sec:sims}.

\input{scalreltable_new}

We note that our framework is set up to perform a full cosmological
analysis. The weak lensing observables are the shear profile ($g_+$ as a function of
angular separation from the cluster center) and $N(z)$, the redshift distribution of the shear source galaxies.  These observables are cosmology independent, as are the SZE signal to noise and redshift for each cluster. However, the likelihood described in equation (\ref{eq:13}) includes cosmological dependencies of the cluster distances and the underlying halo mass function.  All this is built in so that the likelihood can be employed within a full cosmological analysis context.  For the
current work we choose to fix our cosmological parameters to values obtained from
\textit{Planck} and leave the full cosmological analysis to separate work that includes a
larger sample of SPT selected clusters and a subsample that have WL information \citep{bocquet18}.

\subsection{Results for $\zeta-M_{500}$ scaling relation}
\label{sec:results}

\begin{figure*}
\includegraphics[width=\textwidth]{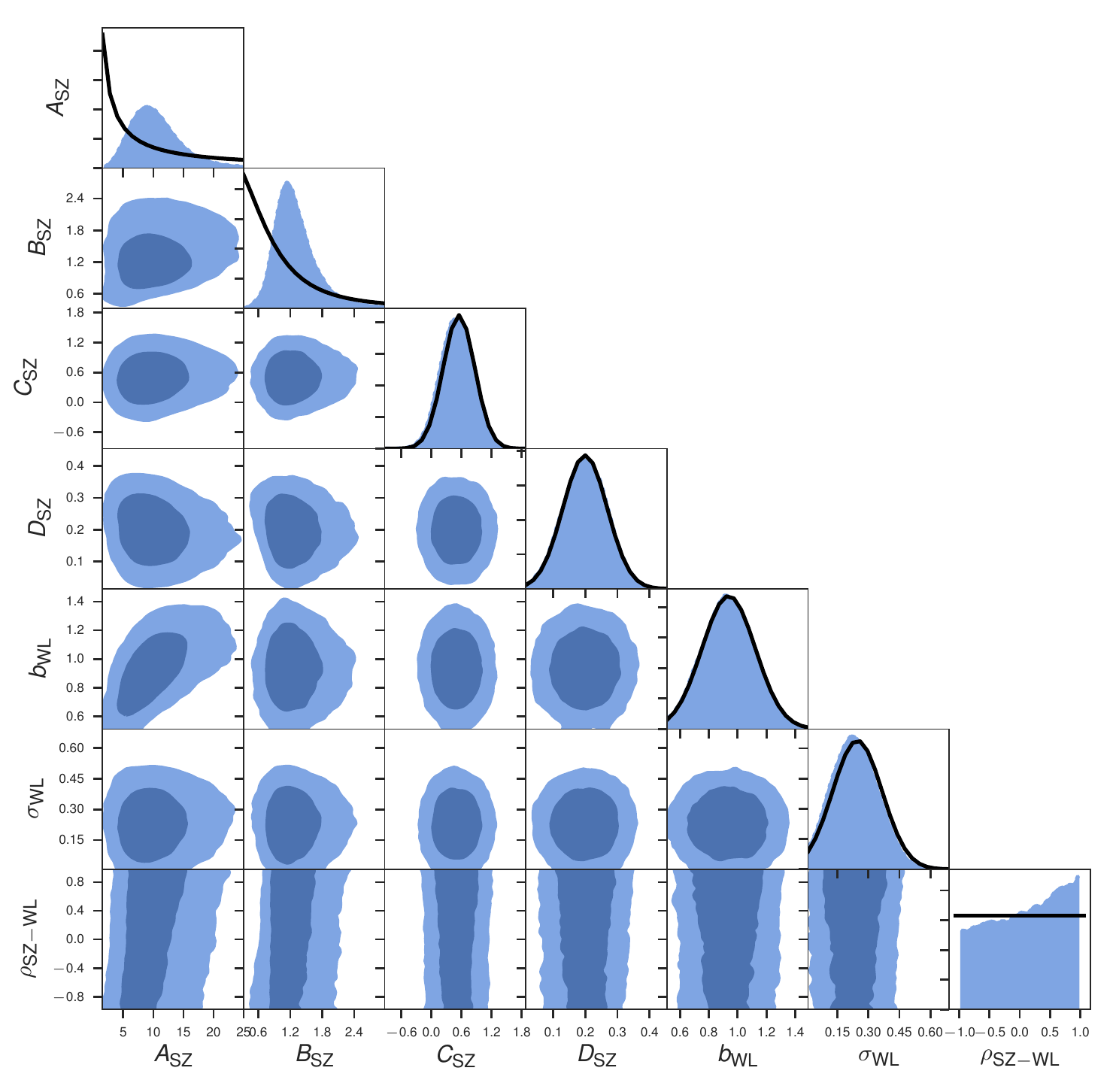}
\vskip-0.2in\caption{Scaling relation parameter constraints derived from our DES-SV WL analysis. Priors (see Section~\ref{sec:priors}) are shown with black lines, where priors for $C_\mathrm{SZ}$ and $D_\mathrm{SZ}$ come from \citet{deHaan16}, and $b_\mathrm{WL}$ and $\sigma_\mathrm{WL}$ arise from our analysis of simulations. Parameters $A_\mathrm{SZ}$, $B_\mathrm{SZ}$ and $\rho_\mathrm{SZ-WL}$ are given broad, uninformative priors and are thus constrained only by WL data. For $A_\mathrm{SZ}$ we find higher values than expected, though still consistent with most previous analyses. Our data prefer an approximately self-similar value for $B_\mathrm{SZ}$, although the uncertainties are large. The data provide no evidence for a correlation between the intrinsic scatter in the SZE-mass and WL-mass scaling relations.}
\label{m500xi}
\end{figure*}

Figure~\ref{m500xi} shows the fully marginalized and joint parameter posterior
distributions from our fit using a recent mass--concentration relation
\citep{DK15} and an uninformative prior on the mass slope
$B_\mathrm{SZ}$. Parameter priors are shown as black solid lines. The
corresponding mean values and the shortest 68\% credible region for each
parameter are presented in Table~\ref{tab:ScalRel}, along with the priors and
literature values from previous SPT studies.

We find $A_\mathrm{SZ}=12.0_{-6.7}^{+2.6}$ when using an informative prior on
the mass slope $B_\mathrm{SZ}$. The probability distribution of
$\ln{A_\mathrm{SZ}}$ is close to Gaussian and there is a tail to high
values. The mean is therefore higher than the mode (9.0) of the
distribution. The correlation coefficient $\rho_\mathrm{SZ-WL}$ is
unconstrained by our data. For the remaining parameters we recover the prior
values.

Because our sample spans a broad range in observable ($\xi=4.5$ to $\xi=42.4$;
and therefore mass), we expect to be able to constrain the mass slope
$B_\mathrm{SZ}$. In the next step we therefore remove the informative prior on
$B_\mathrm{SZ}$, and recover a value of
$B_\mathrm{SZ}=1.30_{-0.44}^{+0.22}$. This value is in agreement with but
somewhat smaller than results from most previous studies
\citep{bocquet15,deHaan16}. Additionally, the normalization shifts down:
$A_\mathrm{SZ}=10.8_{-5.2}^{+2.3}$. This small shift may be caused by a
degeneracy between the parameters, which also explains why the marginalized
uncertainties of $A_\mathrm{SZ}$ decrease. The total posterior volume
increases but rotates in a way that decreases the marginalized uncertainty on
$A_\mathrm{SZ}$ at the expense of increased uncertainty on
$B_\mathrm{SZ}$. The other parameters and parameter uncertainties are
essentially unchanged in comparison to the run with the $B_\mathrm{SZ}$ prior.


We expect significantly tighter constraints on both $A_\mathrm{SZ}$ and $B_\mathrm{SZ}$
with the analysis of the full SPT cluster sample with the DES main survey data. Better
knowledge of the redshift evolution of the SPT mass--observable relation requires
combination with deeper (space-based) data \citep[e.g.,][]{schrabback18}.

\begin{figure*}
\includegraphics[width=\textwidth]{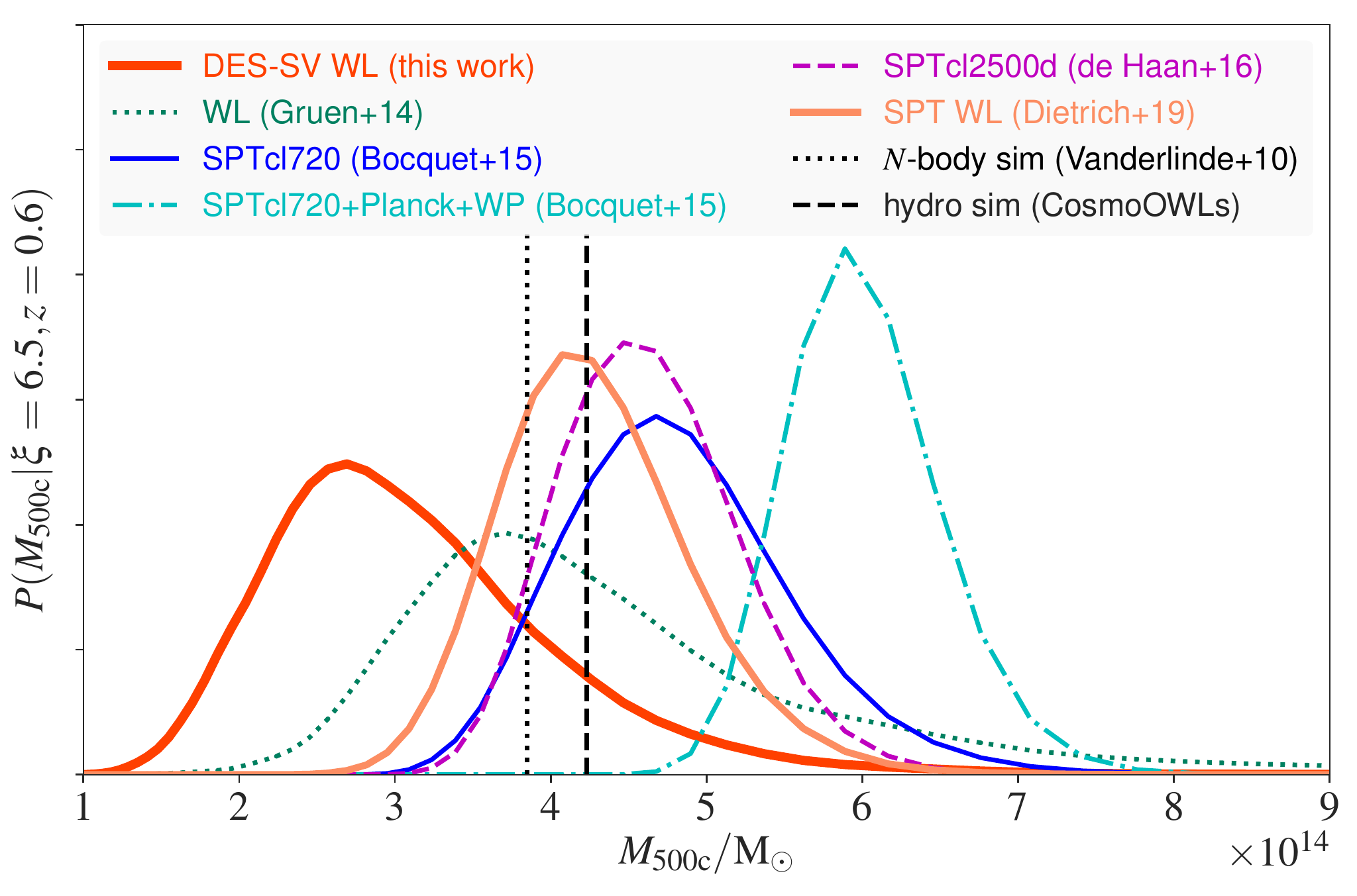}
\vskip-0.20in
\caption{The implications of the best fit SZE mass scaling relation
  expressed as the posterior distribution in mass $M_{500}$ of 
  a typical cluster in the SPT sample ($\xi=6.5$, $z=0.6$).  
  The width of the posterior distribution reflects the 
  parameter uncertainties reported in Table~\ref{tab:ScalRel} and does not include 
  intrinsic scatter or measurement noise on the cluster SZE signature.
  Shown are
  comparable constraints for several different studies as detailed in the text (see also
  Table~\ref{tab:ScalRel}). The vertical lines correspond to the predictions from
  simulations in \citet[dotted line]{vdl10} and the cosmo-OWLS simulation \citep[dashed
  line]{lebrun2014}. As can be seen, previous SPT cosmology analyses recovered higher
  masses than inferred from the WL calibration in this work. When including external
  cosmological parameter priors from CMB anisotropy based analyses \citep{planck13-16},
  even higher masses are preferred.}
\label{Fig:masscomp}
\end{figure*}

\subsection{Comparison to previous results}
\label{sec:discussion}

We now compare our results to SPT mass calibrations in the literature (see Table~\ref{tab:ScalRel}), simulations and  abundance-matching based masses. Additionally, we compare to the shear and magnification studies of smaller SPT selected samples presented in \citet{g1} and \citet{chiu16b}, respectively. 

On the simulation side, the dark matter-only simulations of \citet{vdl10} ($A_\mathrm{SZ}=6.01$ , $B_\mathrm{SZ}=1.31$) and the recent hydrodynamical Cosmo-OWLs simulations \citep[$A_\mathrm{SZ}=5.38$ , $B_\mathrm{SZ}=1.34$;][]{lebrun2014} agree with each other, and are also in agreement with our results given the larger error bars. The smaller value of $B_\mathrm{SZ}$ found when leaving this parameter free is also favoured by \cite{lebrun2014}.

Our measurement is consistent with the clusters-only constraints presented in the latest SPT cosmology analyses \citep{bocquet15,deHaan16}. These studies employ a joint mass calibration and cosmology analysis using mass calibration information from velocity dispersions and X-ray mass proxies. The agreement improves when $B_\mathrm{SZ}$ is left free, though the shift in this parameter from the result of \cite{deHaan16} used as our prior is a promising target for further investigation. \citet{bocquet15} and previous SPT studies recovered a slightly larger value of $D_\mathrm{SZ}$, which is anti-correlated with $A_\mathrm{SZ}$. We therefore attribute a part of the shift in $A_\mathrm{SZ}$ to the use of an updated prior on $D_\mathrm{SZ}$. When including external cosmological priors in a joint mass calibration and cosmological analysis, the external priors--- especially those from CMB measurements--- dominate the cluster mass scale normalization parameter $A_\mathrm{SZ}$ \citep{bocquet15}. This can be seen in the clear shifts of $A_\mathrm{SZ}$ to values below 4, implying masses that are significantly higher than those from this analysis. 

Our normalization of the mass--$\zeta$ relation is also consistent with the recent
weak-lensing calibration of the SPT cluster sample using pointed follow-up observations
\citep{dietrich19}. This approach by the SPT team is similar to ours; it uses the same
calibration on $N$-body simulations and a more recent version of the scaling relations
software employed with a more sophisticated model for the various sources of
weak-lensing scatter. The data sets and the shear catalog creation are, however,
completely independent. While the direct weak-lensing mass calibration of the SPT
cluster sample of \citet{dietrich19} is more in line with expectations from simulations,
velocity dispersion based mass-calibration, and self-calibration of the mass--$\zeta$
relation, the disagreement with our result is not significant, as we will discuss below.

In a previous WL shear analysis of five SPT selected clusters, the WL and SZE based masses were compared \citep{high12}. The mass estimates were in good agreement, with $\langle M_\mathrm{WL}/M_\mathrm{SZ}\rangle=1.07\pm0.18$. We note that the employed SZE masses were from an earlier SPT cluster cosmology analysis \citep{rch13}, and that they were on average about 35\% lower than the masses reported in the analysis of the full SPT-SZ sample \citep{deHaan16}. 

\citet{g1} used WL shear to analyse a sample of SZE-detected clusters, five of which are selected by SPT. The authors constrain the normalization and slope, $A_\mathrm{SZ}=6.0_{-1.8}^{+1.9}$($7.6_{-2.6}^{+3.0}$) and $B_\mathrm{SZ}=1.25_{-0.28}^{+0.36}$($1.02_{-0.68}^{+0.62}$ for a single-halo fit (multi-halo fit, incl. neighbours), when fixing $C_\mathrm{SZ}=0.83$. These values are in agreement with our work, with $C_\mathrm{SZ}$ about $1\sigma$ above the prior and the reported value of \citet{deHaan16}.

Fig.~\ref{Fig:masscomp} illustrates the difference in mass for a hypothetical cluster
with detection signifance $\xi=6.5$ at redshift $z=0.6$. The posterior probabilities for
the cluster mass are calculated by inverting the scaling relation and including the
constraints on $A_\mathrm{SZ}$ and $B_\mathrm{SZ}$. Because this cluster is at the pivot
redshift for our scaling relation, this comparison is insensitive to any difference in
redshift evolution. We neglect the effect of the intrinsic scatter $D_\mathrm{SZ}$ in
this plot, making no correction for the Eddington bias. The bias correction would be
very similar for all sets of constraints we present and thus would have little impact on
the relative differences presented here.

Following \citet{bocquet15} (their Section 5.2), we now calculate the
significance of the differences between our baseline measurement and results
from previous studies. We draw samples from the $P(M)$ for our example
cluster, and calculate the distances $\delta$ of pairs of sampled points. We
estimate $P_{\delta}$, and integrate over the part of the distribution with
$P_{\delta}<P_{\delta}(0)$. In the last step we convert this probability into
a significance assuming a normal distribution. Our result is consistent with
\citet{g1} at the $0.8\sigma$ level, as well as with with previous SPT mass
calibrations \citep{bocquet15, deHaan16} when only clusters are considered
($1.7\sigma$), and the weak lensing calibration of the SPT cluster sample
\citep[$1.4\sigma$,][]{dietrich19}. If one considers the results from
\citet{bocquet15} including additional primary CMB constraints from Planck,
there is tension at the $2.6\sigma$ level.

The use of a larger sample of SPT selected galaxy clusters with DES data will improve constraints on both $A_\mathrm{SZ}$ and $B_\mathrm{SZ}$. At the same time, a complimentary analysis using number count magnification may allow us to further test the stability of WL based mass estimates and our control of systematics. For example, a first magnification study of 19 SPT selected clusters with Megacam data (three clusters overlap with this work) presented in \citet{chiu16b} found a ratio of WL masses to SZE masses of $0.83\pm0.24$, in statistical agreement with both earlier SPT work and this analysis. 

Interestingly, WL derived mass estimates for SPT selected clusters prefer lower values than those from non-WL calibrations, although this preference is not statistically significant given the currently large uncertainties. Given that this is true both for magnification and shear studies from different WL observations, this likely cannot be explained by unknown systematics alone. Larger sample sizes of ongoing WL campaigns are needed to further explore this issue.

If a statistically significant tension between cluster masses calibrated with and without including Planck CMB, BAO and SNe data persists, it may be a hint for new physics. Tensions between CMB anisotropy constraints and constraints from growth-based probes in the context of a $\Lambda$CDM framework can be ameliorated by massive neutrinos or theories of modified gravity. At the same time, possible unknown systematics have to be controlled. We note the disagreement between Planck CMB and cluster cosmology constraints, which may be ameliorated by larger cluster masses (implying a larger bias in their hydrostatic mass estimates). Additionally, the recovered mass from WL also depends on cosmological parameters (especially $h$ and $\Omega_M$). Because the last effect is relatively weak and smaller than the typical precision of current and past analyses, we neglected it in this discussion.

In contrast, a calibration of the optical richness--mass relation through a stacked weak lensing analysis shows good agreement between WL and SZE calibration \citep{saro15,melchior17}.

\section{Conclusions}
\label{sec:conclusions}

In this work we use shear and photo-$z$ catalogs obtained from Science
Verification data taken prior to the start of DES to constrain the masses of
SPT SZE selected clusters of galaxies. The DES catalogs span $139\deg^2$ after
masking and cuts and overlap with 33 SPT selected galaxy clusters above an SPT
SZE significance $\xi>4.5$ and redshifts extending to $z_\mathrm{l}=0.8$.

We first use photo-z's to select the background source galaxies for our WL
study, and then perform a number of cluster-lensing specific tests to further
validate our catalogs. These include examining the shear profiles for the
cross-component, demonstrating that these profiles are consistent with the
expected null signal. We also probe for contamination from cluster galaxies,
using two independent methods to show that there is no measurable
contamination. We examine the dependence of the implied surface mass
overdensity as a function of source properties such as redshift and size,
showing good consistency among all subsamples tested. We demonstrate good
agreement between the two shape catalogs derived using \textsc{ngmix} and
\textsc{im3shape}, though the latter shows lower source number densities,
because it was applied only to $r$-band images.

We then use these validated catalogs to carry out a joint fit of the SZE
mass observable relation, which is described by four parameters
(equation~(\ref{eq:SZmor})). In this process we characterize systematic
biases and intrinsic scatter in WL mass estimates by applying our mass profile
fitting and mass estimation to simulated clusters. We incorporate these
systematics and scatter in our analysis by introducing a WL mass to true mass scaling relation with a free proportionality constant and log-normal scatter (equation~(\ref{eq:wlbias})).

Due to relatively shallow data compared to deeper, pointed WL observations the uncertainties on the masses of individual clusters are relatively large. The availability of shear profiles for the sample of 33 clusters above $z_\mathrm{cl}>0.25$, however, allows one to constrain the $\zeta-M$ relation. For this task we employed an extension of the code developed previously for the cosmological analysis and mass calibration of SPT selected galaxy clusters \citep{bocquet15}. As inputs we use the tangential shear profiles and source redshift distributions, which are direct observables with no cosmological dependence. This approach allows us to self-consistently fit for cosmological and scaling relation parameters. 

For convenience, in this initial study we adopt a flat $\Lambda$CDM cosmology with $\Omega_\mathrm{m}=0.3089$ and $h$=0.6774, as motivated by the latest Planck cosmology analysis \citep{planck2015}. We present parameter constraints on the $\zeta-M$ relation normalization and mass slope. Given the large statistical uncertainties in our shear profiles, we do not expect that marginalizing over the allowed cosmological parameter space consistent with the joint Planck and external dataset analysis would have a significant impact on the scaling relation parameter constraints we derive.

Comparison to earlier SZE mass calibration and cosmology analyses
\citep{bocquet15,deHaan16,dietrich19} shows that our recovered masses are
lower by $\approx 40\%$, but still consistent given the large error bars (see
Figure~\ref{Fig:masscomp}). $A_\mathrm{SZ}$ is insignificantly higher when
adopting a prior on the mass slope $B_\mathrm{SZ}$ from the cosmological
analysis of \citet{deHaan16}. When left free, the recovered mass slope is
shallower than the posterior from \citet{deHaan16}, preferring values closer
to the self-similar expectations for the $\zeta-M$ relation; given the large
uncertainties, the two different slopes are consistent at the 1.2$\sigma$
level.

Our results are in mild tension (at the $2.6\sigma$ level) with the higher cluster masses preferred by primary CMB constraints from Planck \citep{bocquet15}.

The analysis presented in this work has been blinded by multiplying the
overall shear by an unknown factor to avoid observer biases. As mentioned
before, however, in the process of internal collaboration review some
additional tests were requested and have been carried out after unblinding.

This work differs in a number of aspects from the first study of cluster WL using SV data \citep{m1}. That analyses focused on four very massive clusters in a narrow redshift range ($\approx 0.3$--$0.4$) and used a $\Delta \Sigma$ profile as the only ingredient for fitting cluster masses. They used a different photo-$z$ code, which gave only point estimates, an older implementation of \textsc{im3shape} was run on the coadd images and Gaussian errors were adopted.  It also differs from a stacked cluster lensing analysis as presented in \cite{melchior17}, because it uses individual shear profiles and a different treatment of systematics.

The main five-year DES survey will provide full coverage of the SPT footprint at depths somewhat deeper than the data we have used from the SV area. There are 433 confirmed SPT clusters below our redshift limit of $z_\mathrm{cl}<0.8$, that have been imaged by the full survey. A simple scaling with the number of lenses suggests fractional errors of 11.4\% and 8.7\% on $A_\mathrm{SZ}$ and $B_\mathrm{SZ}$ when constraining both parameters simultaneously. To make use of the improved statistical power, further improvement on controlling systematics (see Table~\ref{tab:sysl}) is crucial. This will also impact cluster cosmology which at the moment is limited by our knowledge of the cluster mass scale. To this end, we are proceeding with this broader analysis using the mass calibration method developed for and presented in this paper.

\section*{Acknowledgments}

We acknowledge the support by the DFG Cluster of Excellence Origin and
Structure of the Universe, the Transregio program TR33 ``The Dark Universe''
and the Ludwig-Maximilians-Universit\"at. The analysis presented in this work
benefited from using the computing facilities of the Computational Center for
Particle and Astrophysics (C2PAP) located at the Leibniz-Rechenzentrum (LRZ)
in Munich. DA and TS acknowledge support from the German Federal Ministry of
Economics and Technology (BMWi) provided through DLR under projects
50~OR~1210, 50~OR~1308, 50~OR~1407, and 50~OR~1610. Support for DG was
provided by NASA through Einstein Postdoctoral Fellowship grant number
PF5-160138 awarded by the Chandra X-ray Center, which is operated by the
Smithsonian Astrophysical Observatory for NASA under contract NAS8-03060. AS
is supported by the ERC-StG `ClustersXCosmo', grant agreement 71676. DR is
supported by a NASA Postdoctoral Program Senior Fellowship at NASA's Ames
Research Center, administered by the Universities Space Research Association
under contract with NASA.

The South Pole Telescope is supported by the National Science Foundation through grant ANT-0638937. Partial support is also provided by the NSF Physics Frontier Center grant PHY- 0114422 to the Kavli Institute of Cosmological Physics at the University of Chicago, the Kavli Foundation and the Gordon and Betty Moore Foundation. Galaxy cluster research at Harvard is supported by NSF grant AST- 1009012, and research at SAO is supported in part by NSF grants AST-1009649 and MRI-0723073. The McGill group acknowledges funding from the National Sciences and Engineering Research Council of Canada, Canada Research Chairs Program, and the Canadian Institute for Advanced Research.

We are grateful for the extraordinary contributions of our CTIO colleagues and the DES Camera, Commissioning and Science 
Verification teams in achieving the excellent instrument and telescope conditions that have made this work possible.
The success of this project also relies critically on the expertise and dedication of the DES Data Management organization.

Funding for the DES Projects has been provided by the U.S. Department of Energy, the U.S. National Science Foundation, the Ministry of Science and Education of Spain, 
the Science and Technology Facilities Council of the United Kingdom, the Higher Education Funding Council for England, the National Center for Supercomputing 
Applications at the University of Illinois at Urbana-Champaign, the Kavli Institute of Cosmological Physics at the University of Chicago, Financiadora de Estudos e Projetos, 
Funda{\c c}{\~a}o Carlos Chagas Filho de Amparo {\`a} Pesquisa do Estado do Rio de Janeiro, Conselho Nacional de Desenvolvimento Cient{\'i}fico e Tecnol{\'o}gico and 
the Minist{\'e}rio da Ci{\^e}ncia e Tecnologia, the Deutsche Forschungsgemeinschaft and the Collaborating Institutions in the Dark Energy Survey.

The Collaborating Institutions are Argonne National Laboratory, the University of California at Santa Cruz, the University of Cambridge, Centro de Investigaciones Energeticas, 
Medioambientales y Tecnologicas-Madrid, the University of Chicago, University College London, the DES-Brazil Consortium, the Eidgen{\"o}ssische Technische Hochschule (ETH) Z{\"u}rich, 
Fermi National Accelerator Laboratory, the University of Edinburgh, the University of Illinois at Urbana-Champaign, the Institut de Ciencies de l'Espai (IEEC/CSIC), 
the Institut de Fisica d'Altes Energies, Lawrence Berkeley National Laboratory, the Ludwig-Maximilians Universit{\"a}t and the associated Excellence Cluster Universe, 
the University of Michigan, the National Optical Astronomy Observatory, the University of Nottingham, The Ohio State University, the University of Pennsylvania, the University of Portsmouth, 
SLAC National Accelerator Laboratory, Stanford University, the University of Sussex, and Texas A\&M University.

Facilities: South Pole Telescope, Cerro Tololo Inter-American Observatory's 4 meter Blanco Telescope

\bibliographystyle{mnras}
\bibliography{references}

\bsp


\appendix

\section{Shear profile tests}
\label{sec:shear-profile-tests}
Both shear catalogs we employ have been subjected to an extensive set of tests
described in~\cite{im3_sv}, although as mentioned already
 in Section~\ref{sec:dark-energy-survey}, we have adopted relaxed 
 selection criteria that approximately double the surface density of source galaxies.
 These tests include PSF modelling, $\rho$ statistics
\citep{Rowe_rho} and other tests in the context of various weak lensing
applications like galaxy-galaxy lensing and cosmic shear. To further validate
our shear catalogs within the context of cluster lensing and to justify our
inclusion of fainter and smaller objects, we perform a series of additional
tests. For better statistics we stack our full cluster sample in physical
units. Because several of these tests involve dividing our sources in a
redshift-dependent way, the tangential shear signal generally differs between
different subsamples due to different values of $\beta$. In those cases we
therefore use the (source redshift independent) surface density contrast
$\Delta \Sigma$ defined by
\begin{equation}
\Delta \Sigma = g_+ \times \Sigma_\mathrm{crit}
\end{equation}
instead of $g_+$ for our profile tests. 

In the following subsections we describe results of the following tests:  (1) cross-shear signal, (2) dependence on source signal-to-noise, redshift and size, (3) consistency of the two shear catalogs and (4) stability of $\beta$ distribution to choice of redshift code.

\subsection{Cross shear signal}
\begin{figure}
\includegraphics[width=\columnwidth]{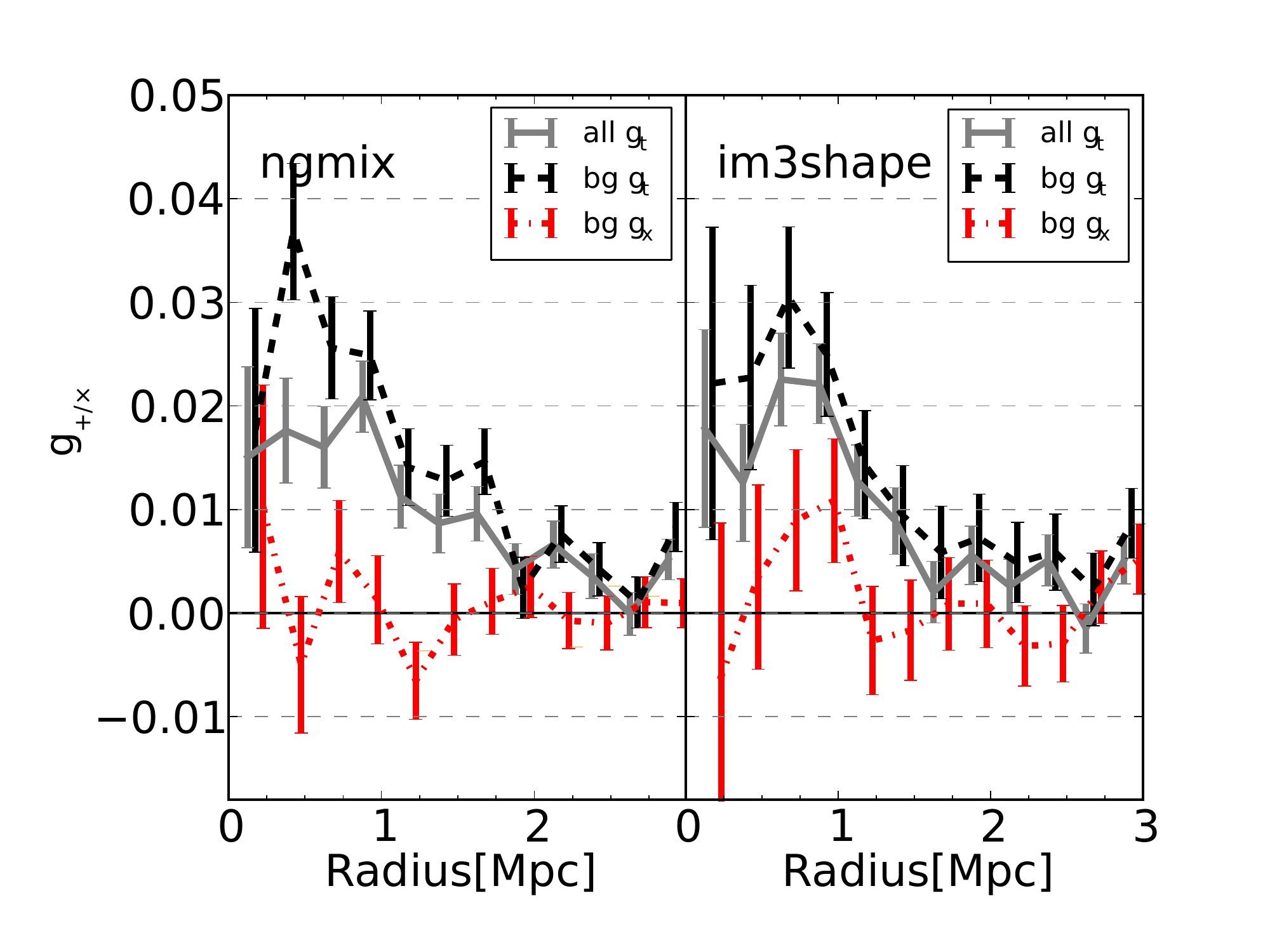}
\vskip-0.2in\caption{Tangential shear signal for the full cluster stack as a function of angular
  cluster-centric distance. The shear signal for all source galaxies, and the background
  galaxies appear in grey and black, respectively, while the cross-shear is in red. The
  cross-shear is consistent with zero, as expected. The error bars are the diagonal
  elements of the covariance matrix computed as in eq.~(\ref{eq:cov}).}
\label{crosstests}
\end{figure}

Figure~\ref{crosstests} contains a plot of several stacked tangential and the
cross shear profiles, where results for \textsc{ngmix} are on the left and
\textsc{im3shape} are on the right. Stacked tangential shear profiles are
shown for the full and background ($z_\mathrm{s}<z_\mathrm{cl}-0.1$) subsample
within 12 linear bins between radii of 0 and 3 Mpc. For each profile we
calculate $\chi^2$ for the null hypothesis of zero shear in the stacked
profiles. We clearly detect the tangential shear signal for the background
sample, obtaining a $\chi^2=167.59$ (73.34) for \textsc{ngmix}
(\textsc{im3shape}). The cross shear has a $\chi^2=7.43$ ($14.17$) for \textsc{ngmix}
(\textsc{im3shape}) with the same binning, which indicates that the data are
consistent with the null hypothesis. These measurements confirm
the validity of our photo-$z$ catalog and that the physical origin of the
shear signal is indeed our lens sample.

\subsection{Dependence on source properties}
\label{sec:depend-source-prop}
\begin{figure}
\includegraphics[width=\columnwidth]{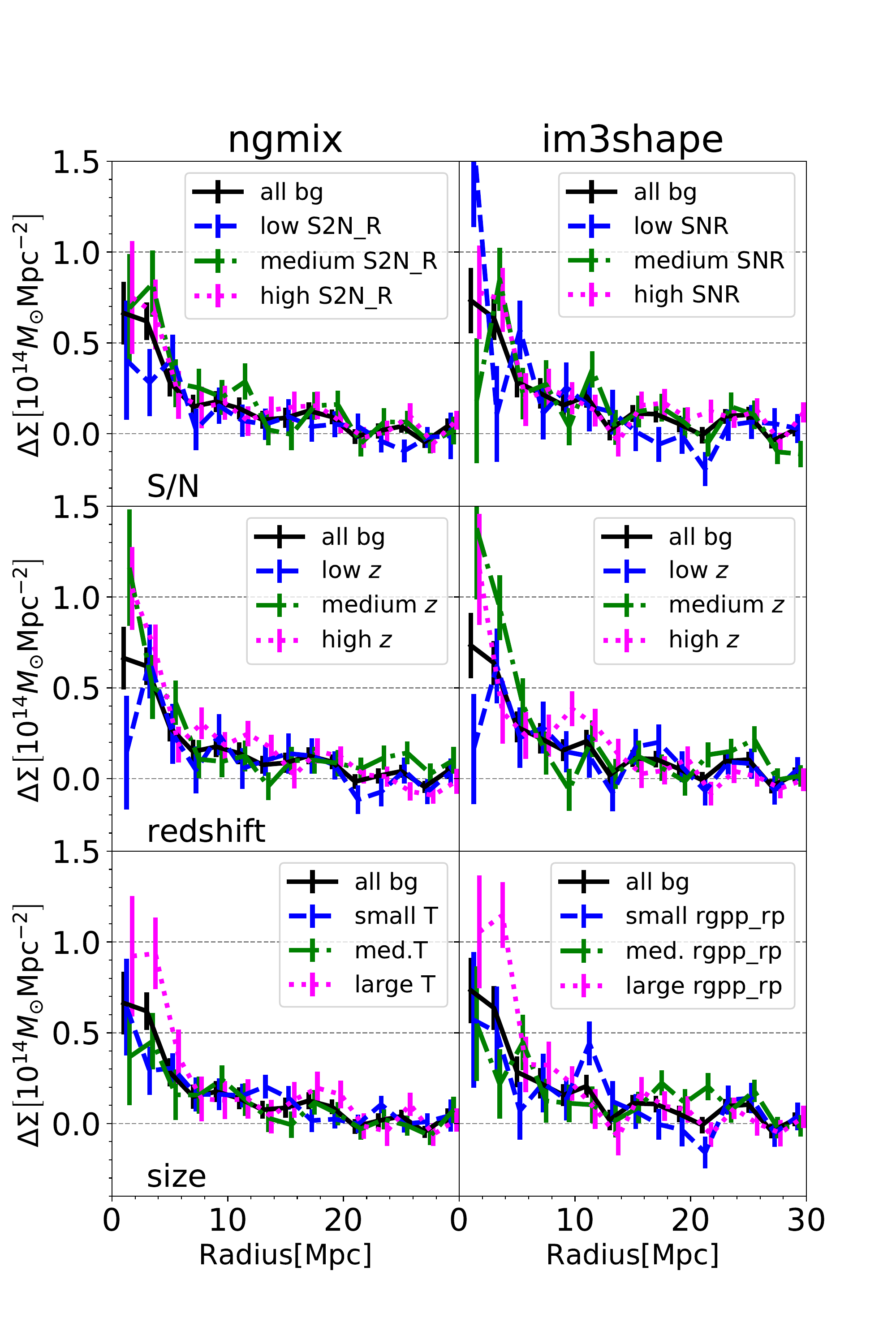}
\vskip-0.3in\caption{Dependence of measured shear signal on source properties in a stack of 28
  $z<0.8$ clusters for \textsc{ngmix} and 37 for \textsc{im3shape}. Each row corresponds
  to a split in one of the following source properties: signal-to-noise, redshift and
  size. We split into equally populated bins for each cluster (instead of a fixed
  boundary for all) to ensure that every source sample sees the same lens sample. Except
  for the case of the large size sources, there is no significant trend of the signal
  with the tested source quantities. The error bars are the standard variation of the
  shear in each radial bin computed via eq.~(\ref{eq:cov}).}
\label{tests}
\end{figure}

Following \cite{m1} we investigate the dependence of our shear signal on
characteristics of the background source population. Figure~\ref{tests} is a
plot of the shear profiles (in angular radius from the cluster center) from
source samples subdivided by signal-to-noise (top), redshift (center) and size
(bottom). \textsc{ngmix} catalogs are on the left, and \textsc{im3shape} on
the right. To construct the profiles, we divide the source sample for each
cluster into three equally populated bins for each quantity of interest. We
take this approach so that each lens contributes the same weight to each of
the three source subsamples. Furthermore, this approach allows us to examine
the impact of the widened selection criteria applied to the shear catalogs
(see Sect.~\ref{sec:dark-energy-survey}). Our selection criteria approximately
double the number of sources by including smaller galaxies and galaxies with
lower SNR. Consequently, our smallest size and lowest SNR bin contain
only sources excluded in the standard cuts; the largest size and highest SNR
bins contain only those galaxies included in the standard cuts, and the
middle bins are approximately equally populated by both kinds of sources. Any
bias caused by the additional sources should then manifest itself as a trend
from small/low bins to large/high bins.

The need to adopt cluster dependent subdivisions of the background sample is
most easily understood in the case of source subsamples divided by redshift,
where clearly the redshift boundaries must shift with the lens redshift. We
note that all three investigated quantities are correlated, with high-$z$
sources typically being smaller and at lower signal-to-noise.

The visual impression within all panels of Figure~\ref{tests} is that all
subsamples are in good agreement. To quantify this, we fit masses using the
shear profiles of each subsample and then compare the consistency of the mass
estimates. For this comparison we fit NFW models to the stacked $\Delta\Sigma$
profiles. We fit $M_{200,\mathrm{c}}$ using the MCMC sampler
\texttt{emcee}\footnote{http://dan.iel.fm/emcee/} for Python~\citep{emcee} and
adopt a flat prior on mass and a log-normal prior on $c$ with
$\sigma_{\ln c}=0.18$. We explicitly allow for negative masses and use the
absolute value of the mass in the mass-concentration relation.

We find excellent agreement within the uncertainties for the source redshift
and SNR subsamples. The only subsample disagreeing with the stacked signal by
more than $1\sigma$ is the \textsc{im3shape} large \texttt{RGPP\_RP} bin at
$1.7\sigma$. Even this subsample, however, agrees at better than $1\sigma$
with the small \texttt{RGPP\_RP} subsample. Furthermore, there is no
consistent trend in mass from small to large size. We thus conclude that there
are no statistically significant trends in inferred cluster mass with object
redshift, SNR, and size, and specifically that the inclusion of additional
objects with low SNR and small size does not lead to a detectable bias in
cluster mass.

\subsection{Consistency between shear catalogs}

Because there are two independent shear catalogs available for DES SV, a comparison between the tangential shear profile between those two is a valuable cross-check. 
We perform this test for the stacked subset of our lens sample that has shape catalogs from both pipelines, and only keep common sources after matching both catalogs.
 
\begin{figure}
\includegraphics[width=9cm]{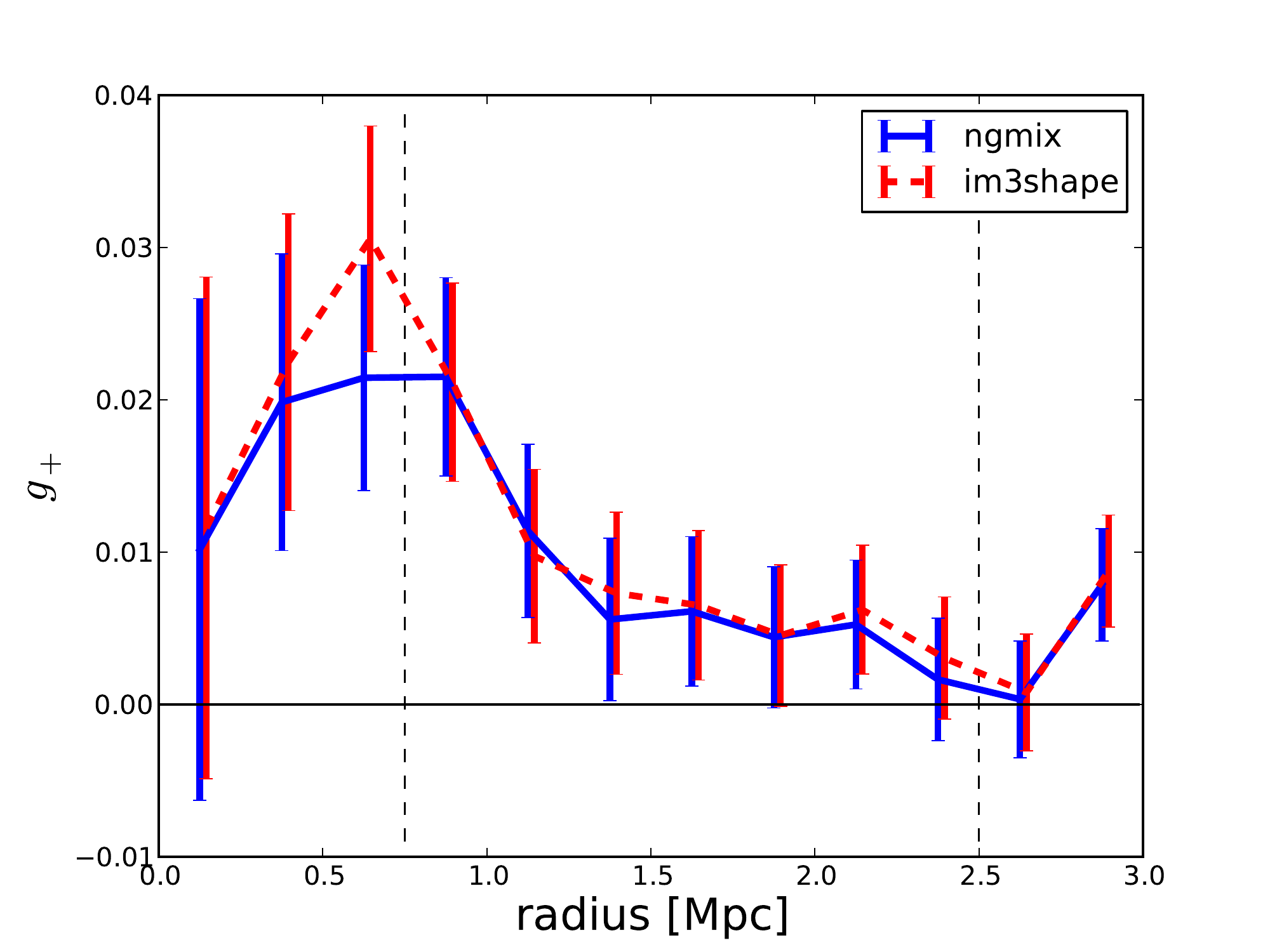} 
\vskip-0.1in\caption{Comparison of the stacked shear signal using \textsc{im3shape} as a second
  independent pipeline for clusters with \textsc{ngmix} coverage. This stack contains 28
  clusters from the SPT-E region, and sources are selected as present in both catalogs
  after all quality cuts. The vertical dashed lines indicate the boundaries of the
  radial fitting range used for the main analysis. The error bars are the diagonal
  elements of the covariance matrix computed as in eq.~(\ref{eq:cov}).}
\label{fig:im3comp}
\end{figure}

Figure~\ref{fig:im3comp} contains the stacked shear profiles of 28 clusters in the
SPT-E field that are covered by both \textsc{ngmix} and \textsc{im3shape}
catalogs, using the same photometric redshifts. For this plot we match both
shape catalogs and only keep common sources that survive all quality
cuts. Additionally, we use the same weights for each galaxy in both
catalogs. In contrast to the previously listed tests, this procedure allows us
to separate out possible redshift estimation problems and focus directly on
the shear measurement.   For further discussion of comparison of the 
\textsc{ngmix} and \textsc{im3shape} shear catalogs, we refer the reader 
to \citet{im3_sv}.

The cuts employed in this work are less strict than those for analyses that
use the full SPT-E footprint. To test the dependence of our result on the
details of our cut, we compare stacked tangential shear profiles for
\textsc{ngmix} using both our standard cuts and the most conservative cuts in
\cite{im3_sv}. The profiles are fully consistent in both cases, although the
signal-to-noise is degraded with the stricter cuts due to lower number density
of source galaxies. Therefore, we believe that no additional bias in incurred
by the relaxed selection criteria. We adopt a Gaussian prior on the
multiplicative bias with a standard deviation of 15\%, based on extrapolating
the behavior of $m$ found in \cite{im3_sv} to the expanded selection.

\subsection{Stability of $\beta$-estimation}
\label{beta_section}
\begin{figure}
\includegraphics[width=9cm]{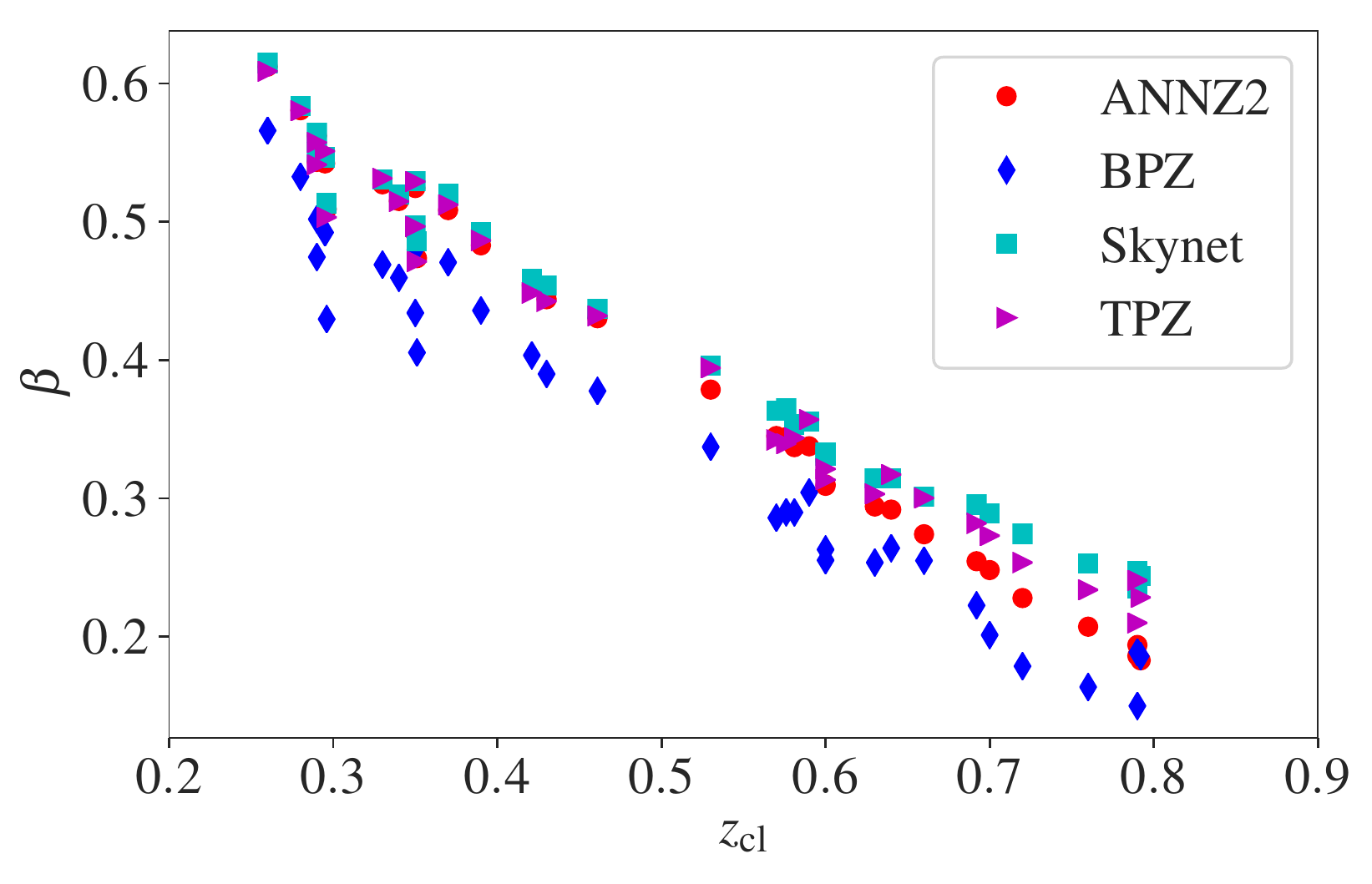}
\vskip-0.1in\caption{Estimated lensing efficiency $\beta$ for our background source
  population for each cluster when employing different redshift codes. The
  different values are plotted against the cluster redshift
  $z_\mathrm{cl}$. The three training-set based methods (\textsc{annz2},
  \textsc{tpz}, and \textsc{skynet} show the best agreement while the template
  based \textsc{bpz} code disagrees most strongly with the fiducial
  \textsc{skynet} algorithm.}
\vskip-0.15in
\label{betaplot}
\end{figure}

We calculate $\beta$ values from source redshift distributions $N(z)$ for each
cluster assuming our standard cosmology (flat $\Lambda$CDM,
$\Omega_\mathrm{m}=0.3089$) for all four different photo-$z$ codes available
within DES. Figure~\ref{betaplot} shows the results for our tests. Overall
all methods show reasonable agreement. Larger discrepancies exist between
the template fitting code \textsc{bpz} and the three training-set based
methods. The mean of the $\beta$ differences is 6.5\% when comparing
\textsc{skynet} and \textsc{bpz}. We consider this as an estimate of the
systematic uncertainties when estimating the lensing efficiencies of the SPT
clusters in DES SV data. This uncertainty in $\beta$ translates into a
systematic mass uncertainty of 9.6\%, which is what we use as the standard
deviation of a Gaussian prior.

To estimate the influence of imperfect knowledge of the lens redshifts, we resample
$D_\mathrm{l}\beta$ 100 times for each cluster from a Gaussian distribution with width
equal to the redshift error of our photometric lens sample. We find a mean ratio of
$1.010 \pm 0.015$ relative to taking the centre of the redshift distribution for our
full sample. Because this is consistent with unity and negligible compared to the other
sources of systematic uncertainty listed in Table~\ref{tab:sysl}, we can safely ignore
uncertainties in the lens redshifts.

\section{Priors on weak lensing mass bias and scatter}
\label{sec:wlpriors}

To obtain the priors on the weak lensing mass bias $b_\text{WL}$ and scatter $\sigma_\text{WL}$ (equation~\ref{eq:wlbias}), we create an ensemble of simulated observations that match the observational properties of a random subset of cluster fields and then apply the same measurement technique as we do to the real data. In general, we are aiming to reconstruct the probability distribution $P(M_{\mathrm{meas}}|M_{\mathrm{true}})$, which can then be included in forward modelling of the cluster sample. However, we simplify the relation as stated above to one log-normal distribution that is the same for all observed cluster fields. Any residuals from such an oversimplification are still insignificant compared to the obtainable statistical precision of our dataset.

To build our simulated observations for one observed cluster field, we start with the
$N$-body simulations from \citet{beckerkravtsov}, with parameters
$\Omega_\mathrm{m} = 0.27$, $\Omega_\mathrm{b} = 0.044$, $\sigma_8 = 0.79$, spectral
index $n=0.95$ and a Hubble constant of $h=0.7$ in units of
100\,km\,s$^{-1}$\,Mpc$^{-1}$. We cut out $400h^{-1}$Mpc long boxes centered on the most
massive 788 halos with $M_{500,c} > 1.5\times10^{14}h^{-1}M_\odot$ from the $z=0.5$
snapshot. Particles are projected to form 2D mass maps that are then used to create
shear maps via Fast Fourier transform. The observed $\langle\beta\rangle$ from a cluster
observation is used to scale the shear and kappa maps appropriately. Random Gaussian
noise is added to the shear map to match the observed shape noise in the
observations. Because in our real observations we fit a 1-D profile, we select an
"observed" cluster center for each simulation map. We assume that the displacement
between the true projected center of the simulated cluster and the "observed" center is
randomly oriented with respect to the underlying structure, a not unreasonable
assumption given the noise sources of SPT observations and the statistical power of this
sample. Center offsets are randomly chosen following the form specified by
\citet{song12b}, a Gaussian distribution with a width dependent on the SPT beam size and
the core radius of the matched filter used to detect the observed cluster. The simulated
1-D profiles are then fit with an NFW model as in the data analysis.

We assume that $P(M_{\mathrm{meas}}|M_{\mathrm{true}})$ follows a log-normal
distribution where $\ln b_{\mathrm{WL}}$ is the mean of the distribution and
$\sigma_{\mathrm{WL}}$ is its width. For the set of simulated fields, we find
the maximum \emph{a posteriori} location for the probability distribution
\begin{equation}
\begin{split}
&P(b_{\mathrm{WL}}, \sigma_{\mathrm{WL}} | \mathrm{mocks}) \propto  \\ 
&\prod_i \int P(b_{\mathrm{WL}}, \sigma_{\mathrm{WL}}
|M_{\mathrm{meas}})P(M_\mathrm{meas} | \mathrm{mock}_i) \mathrm{d}M_{\mathrm{meas}}. 
\end{split}
\end{equation}
Uninformative priors are used for the parameters of interest. Simulated observations are also created and analyzed using the $z=0.25$ snapshot from \citet{beckerkravtsov} as well as the Millennium-XXL simulations \citep{angulo2012}. No significant trends are seen between snapshots or simulations. We also do not see any significant trend with the observational properties of each observed field, including the amount of shape noise or different filter core size. Our final bias number is then the average of $b_{\mathrm{WL}}$ across the random subset of cluster fields targeted for mock up.

We measure $b_\mathrm{WL}=0.936\pm0.04$ and $\sigma_\mathrm{WL}=0.25\pm0.12$ when
employing the $M$--$c$ relation of \citet{DK15}. In practice we add the systematic
uncertainty of the weak-lensing mass to true mass relation to all other sources
systematic errors (see Table~\ref{tab:sysl}) in quadrature and use a Gaussian prior
$b_\mathrm{WL} = 0.936 \pm 0.185$ in our scaling relation analysis.

We can estimate the sensitivity of our analysis to the uncertainty in published
mass--concentration relations by carrying out the NFW fit bias analysis for different
fixed concentrations. We find that the average mass bias at concentrations $c = 5$ and
$c = 3$ is $b_\mathrm{WL} = 0.978$ and $b_\mathrm{WL} = 0.907$, respectively, implying
$\mathrm{d} b_\mathrm{WL} / \mathrm{d} c |_{c=4} = -0.0355$. Using Gaussian error
propagation on eq.~(\ref{eq:wlbias}) we obtain
\begin{equation}
  \label{eq:21}
  \left(\frac{\sigma_M}{M_\mathrm{true}}\right)^2 = \frac{1}{b_\mathrm{WL}^2}
  \left(\frac{\mathrm{d} b_\mathrm{WL}}{\mathrm{d} c}\right)^2 \sigma_c^2\;.
\end{equation}
Because we calibrated the bias resulting from an NFW fit using the $M\text{--}c$
relation of \citet{DK15}, the systematic uncertainty is not given by how well this
relation describes the actual cluster sample, but by how faithfully the simulated
clusters represent true clusters in the Universe. The simulations used in the previous
section are Dark Matter only for a cosmology consistent with WMAP7 \citep{komatsu11} and
thus the question is how much would the concentrations for clusters of the mass and
redshift in our sample be impacted by baryonic effects and the change of cosmological
parameters to ones consistent with \textit{Planck}. \citet{duffy10} constrain baryonic
effects to an upper limit of $10$\,per~cent, with baryons decreasing the concentration
value. \citet{klypin16} show that concentrations are $\sim10$~per~cent larger in
\textit{Planck} cosmologies than in the \textit{WMAP} cosmology assumed in our
calibration of the weak lensing bias $b_\mathrm{WL}$. One could thus expect these
effects to cancel but we conservatively assume an uncertainty of 10~per~cent on the
concentration. Evaluating eq.~(\ref{eq:21}) we set $\sigma_c|_{c=4}=0.4$ and obtain a
mass uncertainty due to the mass--concentation relation of $1.5\%$. This turns out to be
so much smaller than our other systematic uncertainties (cf. Table~\ref{tab:sysl}) that
we can safely ignore it.

\section*{Affiliations}
\input{affiliations.tex}

\label{lastpage}
\end{document}

%% file: authors.tex

\author[Stern et al.]{
\parbox{\textwidth}{
\Large
C.~Stern$^{1,2}$,
J.~P.~Dietrich\thanks{dietrich@usm.lmu.de}$^{1,2}$,
S.~Bocquet$^{3}$,
D.~Applegate$^{4,5}$,
J.~J.~Mohr$^{1,2,6}$,
S.~L.~Bridle$^{7}$,
M.~Carrasco~Kind$^{8,9}$,
D.~Gruen\thanks{Einstein Fellow}$^{10,11}$,
M.~Jarvis$^{12}$,
T.~Kacprzak$^{13}$,
A.~Saro$^{1,14}$,
E.~Sheldon$^{15}$,
M.~A.~Troxel$^{16,17}$,
J.~Zuntz$^{18}$,
B.~A.~Benson$^{19,20,21}$,
R.~Capasso$^{1,2}$,
I.~Chiu$^{1,2}$,
S.~Desai$^{22}$,
D.~Rapetti$^{23,24}$,
C.~L.~Reichardt$^{25}$,
B.~Saliwanchik$^{26}$,
T.~Schrabback$^{5}$,
N.~Gupta$^{1,2}$,
T.~M.~C.~Abbott$^{27}$,
F.~B.~Abdalla$^{28,29}$,
S.~Avila$^{30}$,
E.~Bertin$^{31,32}$,
D.~Brooks$^{28}$,
D.~L.~Burke$^{10,11}$,
A.~Carnero~Rosell$^{33,34}$,
J.~Carretero$^{35}$,
F.~J.~Castander$^{36,37}$,
C.~B.~D'Andrea$^{12}$,
L.~N.~da Costa$^{33,34}$,
C.~Davis$^{10}$,
J.~De~Vicente$^{38}$,
H.~T.~Diehl$^{19}$,
P.~Doel$^{28}$,
J.~Estrada$^{19}$,
A.~E.~Evrard$^{39,40}$,
B.~Flaugher$^{19}$,
P.~Fosalba$^{36,37}$,
J.~Frieman$^{19,21}$,
J.~Garc\'ia-Bellido$^{41}$,
E.~Gaztanaga$^{36,37}$,
R.~A.~Gruendl$^{8,9}$,
J.~Gschwend$^{33,34}$,
G.~Gutierrez$^{19}$,
D.~Hollowood$^{42}$,
T.~Jeltema$^{42}$,
D.~Kirk$^{28}$,
K.~Kuehn$^{43}$,
N.~Kuropatkin$^{19}$,
O.~Lahav$^{28}$,
M.~Lima$^{44,33}$,
M.~A.~G.~Maia$^{33,34}$,
M.~March$^{12}$,
P.~Melchior$^{45}$,
F.~Menanteau$^{8,9}$,
R.~Miquel$^{46,35}$,
A.~A.~Plazas$^{47}$,
A.~K.~Romer$^{48}$,
E.~Sanchez$^{38}$,
R.~Schindler$^{11}$,
M.~Schubnell$^{40}$,
I.~Sevilla-Noarbe$^{38}$,
M.~Smith$^{49}$,
R.~C.~Smith$^{27}$,
F.~Sobreira$^{50,33}$,
E.~Suchyta$^{51}$,
M.~E.~C.~Swanson$^{9}$,
G.~Tarle$^{40}$,
A.~R.~Walker$^{27}$
\begin{center} (DES \& SPT Collaborations) \end{center}
}}

%% file: spt_clusters_2500d.tex

\begin{table*}
 \caption{Lens sample used. From left, we list the name, sky position, SZE
   significance, detection scale $\theta_\mathrm{c}$, SZE
   $M_{500,\mathrm{SZ}}$, redshift (where "(s)" denotes spectroscopic
   redshift), DES Field (SNE= ELAIS supernova field) and whether
   \textsc{ngmix} catalog is available.  The $\dagger$ marks clusters
   presented in \citet{saro15}} %
 \centering
 \begin{tabular}{lclccrccc}
 \hline\hline
  & R.A. & Dec. & & $\Theta_\mathrm{c}$ & $M_{500,\mathrm{SZ}}$ &  &  &  \\
SPT ID & [deg] & [deg] & $\xi$ & [arcmin] & [$10^{14}\, \mathrm{M}_\odot$] & Redshift & DES Field & \textsc{ngmix} \\
  \hline
  \hline
SPT-CL J0040$-$4407 &10.2048 &$-$44.1329 & 19.34 & 0.50 & $10.24\pm1.56$ & $0.350$(s) & SNE\\[3pt]
SPT-CL J0041$-$4428 &10.2513 &$-$44.4785& 8.84 & 0.50 & $5.83\pm1.01$ & $0.33\pm 0.02$ & SNE\\[3pt]
SPT-CL J0107$-$4855 &16.8857 &$-$48.9171& 4.51 & 0.25 & $2.48 \pm 0.81$ & $0.60\pm 0.03$ & El Gordo & \\[3pt]

SPT-CL J0412$-$5106 &63.2297 &$-$51.1098 & 5.15 & 0.25 & $3.42\pm0.84$& $0.28\pm 0.04$ & SPT-E & \checkmark \\[3pt]
SPT-CL J0417$-$4748 &64.3451 &$-$47.8139 & 14.24 & 0.25 &$ 7.41\pm1.15$& $0.581$(s) & SPT-E & \checkmark \\[3pt]
SPT-CL J0422$-$4608 &65.7490 &$-$46.1436 & 5.05 & 0.50 &$ 2.90\pm0.75$& $0.70\pm 0.03$ & SPT-E & \checkmark \\[3pt]
SPT-CL J0422$-$5140 &65.5923 &$-$51.6755 & 5.86 & 1.00 & $3.57\pm0.77$ & $0.59\pm 0.03$& SPT-E & \checkmark \\[3pt]
SPT-CL J0426$-$5455 &66.5199 &$-$54.9197 & 8.85 & 0.50 & $5.17\pm0.90$& $0.63\pm 0.03$& SPT-E & \checkmark \\[3pt] 

SPT-CL J0428$-$6049 &67.0305 &$-$60.8292 & 5.11 & 1.25 & $3.04\pm0.78$ & $0.64\pm 0.03$& SPT-E & \checkmark \\[3pt]
SPT-CL J0429$-$5233 &67.4315 &$-$52.5609 & 4.56 & 0.75 & $2.75\pm0.77$& $0.53 \pm 0.03$& SPT-E & \checkmark \\[3pt]
SPT-CL J0433$-$5630 &68.2541 &$-$56.5025 & 5.32 & 1.75 &$3.13\pm0.76$& $0.692$(s)& SPT-E & \checkmark \\[3pt]
SPT-CL J0437$-$5307 &69.2599 &$-$53.1206& 4.52 & 0.25 & $3.20\pm0.80\dagger$ & $0.29 \pm 0.02 \dagger$& SPT-E & \checkmark \\[3pt]
SPT-CL J0438$-$5419 &69.5749 &$-$54.3212& 22.88 & 0.50 & $10.80\pm1.62$& $0.421$(s)& SPT-E & \checkmark \\[3pt]

SPT-CL J0439$-$4600 &69.8087 &$-$46.0142 & 8.28 & 0.25 & $5.29\pm0.94$& $0.34\pm 0.04$ & SPT-E & \checkmark \\[3pt]
SPT-CL J0439$-$5330 &69.9290 &$-$53.5038 & 5.61 & 0.75 & $3.59\pm0.80$& $0.43 \pm 0.04$& SPT-E & \checkmark \\[3pt]
SPT-CL J0440$-$4657 &70.2307 &$-$46.9654 & 7.13 & 1.25 & $4.63\pm0.89$ & $0.35 \pm 0.04$& SPT-E & \checkmark \\[3pt]
SPT-CL J0441$-$4855 &70.4511 &$-$48.9190& 8.56 & 0.50 & $4.74\pm0.83$ & $0.79 \pm 0.04$& SPT-E & \checkmark \\[3pt]
SPT-CL J0444$-$4352 &71.1683 &$-$43.8735 & 5.01 & 1.50 & $3.11\pm0.82$ & $0.57 \pm 0.03$& SPT-E & \checkmark \\[3pt]

SPT-CL J0447$-$5055 &71.8445 &$-$50.9227 & 5.96 & 0.25 & $3.87\pm0.82$&$0.39 \pm 0.05$& SPT-E & \checkmark \\[3pt]
SPT-CL J0449$-$4901 &72.2742 &$-$49.0246 & 8.91 & 0.50 &$4.90\pm0.85$ & $0.792$(s)& SPT-E & \checkmark \\[3pt]
SPT-CL J0452$-$4806 &73.0034 &$-$48.1102 & 4.52 & 0.50 & $2.87\pm0.81$& $0.37 \pm 0.04$& SPT-E & \checkmark \\[3pt]
SPT-CL J0456$-$5623 &74.1753 &$-$56.3855 & 4.60 & 0.25 &$2.68\pm0.75$ & $0.66 \pm 0.03$& SPT-E & \checkmark \\[3pt]

SPT-CL J0500$-$4551 &75.2108 &$-$45.8564 & 4.51 & 0.75 & $3.60\pm0.91\dagger$ & $0.26 \pm 0.01 \dagger$& SPT-E & \checkmark \\[3pt]
SPT-CL J0502$-$6048  & 75.7240  & $-$60.810 &  4.69 & 0.25 & $3.03\pm0.76\dagger$ & $ 0.79 \pm 0.02  \dagger$ & SPT-E & \checkmark \\[3pt]
SPT-CL J0509$-$5342 &77.3374 &$-$53.7053 & 8.50 & 0.75 & $5.06\pm0.89$& $0.461$(s)& SPT-E & \checkmark \\[3pt]

SPT-CL J0516$-$5430 &79.1513 &$-$54.5108 & 12.41 & 1.50 & $7.10\pm1.14$& $0.295$(s)& SPT-E & \checkmark \\[3pt]
SPT-CL J0529$-$6051 &82.3493 &$-$60.8578& 5.58 & 0.50 & $3.39\pm0.78$& $0.72 \pm 0.06$& SPT-E & \checkmark \\[3pt]
SPT-CL J0534$-$5937 &83.6082 &$-$59.6257 & 4.74 & 0.25 & $2.75\pm0.75$& $0.576$(s)& SPT-E & \checkmark \\[3pt]
SPT-CL J0540$-$5744 &85.0043 &$-$57.7405 & 6.74 & 0.25 & $3.76\pm0.74$& $0.76 \pm 0.03$& SPT-E & \checkmark \\[3pt]
SPT-CL J0655$-$5541 &103.9137 &$-$55.6931 & 5.64 & 1.00 & $3.98\pm0.88$& $0.29 \pm 0.04$& Bullet & \\[3pt]

SPT-CL J0658$-$5556 &104.6317 &$-$55.9465 & 39.05 & 1.25 & $16.86\pm2.49$& $0.296$(s)& Bullet & \\[3pt]
SPT-CL J2248$-$4431 &342.1907 &$-$44.5269 & 42.36 & 0.75 & $17.27\pm2.54$& $0.351$(s)&RXJ2248 & \\[3pt]
SPT-CL J2249$-$4442 &342.4069 &$-$44.7158 & 5.11 & 0.25 & $3.18\pm0.81$& $0.60 \pm 0.03$&RXJ2248 &\\[3pt]
\hline\hline
\end{tabular}
 \label{tab:SPTclusters}
\end{table*}

%% file: spt_clusters_2500d_wl.tex

\begin{table}
	\centering
\caption{Weak Lensing Information for each cluster, where
$N_\mathrm{gal}$ denotes the number of background galaxies used for fitting, and $N_\mathrm{bin}$ is the number of radial bins.  These quantities are shown both for the \textsc{ngmix} and the \textsc{im3shape} catalogs. The last columns contains the median $r$-band seeing $\theta_\mathrm{psf}$ within a 10~arcmin aperture centered on each cluster.}
 \begin{tabular}{lccccc}
 \hline\hline
SPT ID & $N_\mathrm{gal}^\mathrm{NG}$ & $N_\mathrm{bin}$  & $N_\mathrm{gal}^\mathrm{im3}$ & $N_\mathrm{bin}$ & $\theta_\mathrm{psf}$\\ 
\hline \hline
SPT-CL J0040$-$4407 & \ldots & \ldots & 634 & 5 & 1\farcs25 \\[3pt]
SPT-CL J0041$-$4428 &  \ldots & \ldots & 351 & 5 & 1\farcs26 \\[3pt]
SPT-CL J0107$-$4855 &  \ldots & \ldots & 200 & 5 & 1\farcs15 \\[3pt]

SPT-CL J0412$-$5106 & 2074 & 10 & \ldots & \ldots & 1\farcs23 \\[3pt]
SPT-CL J0417$-$4748 &  385 &  5 & \ldots & \ldots & 1\farcs18 \\[3pt]
SPT-CL J0422$-$4608 &  266 &  5 & \ldots & \ldots & 1\farcs11 \\[3pt]
SPT-CL J0422$-$5140 &  429 &  5 & \ldots & \ldots & 1\farcs18 \\[3pt]
SPT-CL J0426$-$5455 &  238 &  5 & \ldots & \ldots & 1\farcs30 \\[3pt] 

SPT-CL J0428$-$6049 &  518 &  5 & \ldots & \ldots & 1\farcs04 \\[3pt]
SPT-CL J0429$-$5233 &  550 &  5 & \ldots & \ldots & 1\farcs14 \\[3pt]
SPT-CL J0433$-$5630 &  239 &  5 & \ldots & \ldots & 1\farcs24 \\[3pt]
SPT-CL J0437$-$5307 & 2276 & 11 & \ldots & \ldots & 1\farcs18 \\[3pt]
SPT-CL J0438$-$5419 &  961 &  5 & \ldots & \ldots & 1\farcs29 \\[3pt]

SPT-CL J0439$-$4600 & 1608 &  8 & \ldots & \ldots & 1\farcs18 \\[3pt]
SPT-CL J0439$-$5330 &  987 &  5 & \ldots & \ldots & 1\farcs22\\[3pt]
SPT-CL J0440$-$4657 & 2168 & 10 & \ldots & \ldots & 1\farcs16\\[3pt]
SPT-CL J0441$-$4855 &  362 &  5 & \ldots & \ldots & 1\farcs14\\[3pt]
SPT-CL J0444$-$4352 &  408 &  5 & \ldots & \ldots & 1\farcs24\\[3pt]

SPT-CL J0447$-$5055 & 1547 &  7 & \ldots & \ldots & 1\farcs19\\[3pt]
SPT-CL J0449$-$4901 &  420 &  5 & \ldots & \ldots & 1\farcs05\\[3pt]
SPT-CL J0452$-$4806 & 1914 &  9 & \ldots & \ldots & 1\farcs10\\[3pt]
SPT-CL J0456$-$5623 &  420 &  5 & \ldots & \ldots & 1\farcs24\\[3pt]
SPT-CL J0500$-$4551 & 2500 & 12 & \ldots & \ldots & 1\farcs20\\[3pt]

SPT-CL J0502$-$6048 &  336 &  5 & \ldots & \ldots & 1\farcs10\\[3pt]
SPT-CL J0509$-$5342 &  702 &  5 & \ldots & \ldots & 1\farcs23\\[3pt]
SPT-CL J0516$-$5430 & 1541 &  7 & \ldots & \ldots & 1\farcs21\\[3pt]
SPT-CL J0529$-$6051 &  169 &  5 & \ldots & \ldots & 1\farcs23\\[3pt]
SPT-CL J0534$-$5937 &  414 &  5 & \ldots & \ldots & 1\farcs28\\[3pt]

SPT-CL J0540$-$5744 &  174 &  5 & \ldots & \ldots & 1\farcs24\\[3pt]
SPT-CL J0655$-$5541 & \ldots & \ldots & 519 & 5 & 1\farcs06\\[3pt]
SPT-CL J0658$-$5556 & \ldots & \ldots & 691 & 5 & 1\farcs06\\[3pt]
SPT-CL J2248$-$4431 & \ldots & \ldots & 593 & 5 & 1\farcs22\\[3pt]
SPT-CL J2249$-$4442 & \ldots & \ldots & 194 & 5 & 1\farcs17\\[3pt]
\hline\hline
  \end{tabular}
\label{tab:SPTWL}
\end{table}

%% file: sys_table.tex
\begin{table}
  \centering
  \caption{Systematic mass error budget broken down into contributions from
    the source redshift distribution $\beta$, the multiplicative shear bias
    $m$ and the cluster contamination $f_\mathrm{500}$. We additionally
    consider errors due to miscentering, deviations from an NFW profile, as
    calibrated by simulations and parametrized by $b_\mathrm{WL}$.
    References are provided in column 4. The total systematic
    uncertainty consists of the listed effects added in quadrature.}
  \begin{tabular}{lccc}
    \hline\hline
    Systematic & Error & $\Delta M_{500}$ & Reference\\
    \hline \hline
    $\beta$ & 6.5\% & 9.6\% &  \S~\ref{beta_section}, \cite{photoz4WL} \\
    \multirow{2}{*}{$m$} & \multirow{2}{*}{10\%} & \multirow{2}{*}{15\%} & extrapolated from \\
    & & & \cite{im3_sv} \\
    $f_{500}$ & 6.9\% & 3.4\% & \S~\ref{sec:contamination} \\
    $b_\mathrm{WL}$ & 4.0\% & 4.0\% & \S~\ref{sec:sims} \\
    \hline
    Total & & 18.6\% & \\
    \hline\hline
  \end{tabular}
  \label{tab:sysl}
\end{table}

%% file: scalreltable_new.tex
\begin{table*}
	\centering
  \caption{$\zeta$--$M_{500}$ scaling relation parameter constraints and priors for three previous SPT publications as well as this analysis (DES-SV WL Shear).  Constraints are shown for the four SZE-mass relation parameters and the two WL mass-mass relation.  WL results are shown when adopting the mass-concentration relation from \citet{DK15}.  Results are shown with and without a prior on $B_\mathrm{SZ}$.
  }
\begin{tabular}{lcccccc}
\hline\hline
Analysis and Constraints          & $A_\mathrm{SZ}$ & $B_\mathrm{SZ}$ & $C_\mathrm{SZ}$ &$  D_\mathrm{SZ}$ & $b_\mathrm{WL}$ & $\sigma_\mathrm{WL}$ \\
\hline\hline
\citet{bleem15} fixed parameters  &        $4.14$ &         $1.44$ &         $0.59$ &         $0.22$ &            \ldots &                 \ldots \\
\citet{bocquet15} $\mathrm{SPT}_\mathrm{CL}$+$Y_x$+$\sigma_v$ 
                                  &$4.7_{-1.2}^{+0.8}$& $1.58\pm0.12$&    $0.91\pm0.35$& $0.26\pm 0.10$ &            \ldots &                 \ldots\\
\ \     +Planck+WP+BAO+SNe        & $3.2\pm0.3$    & $1.49\pm0.11$&    $0.49\pm0.22$& $0.26\pm 0.05$  &             \ldots&                \ldots\\
	    
\citet{deHaan16} $\mathrm{SPT}_\mathrm{CL}$+$Y_x$                   
                                  & $4.8\pm0.9$    & $1.67\pm0.08$ & $0.55\pm0.32$& $0.20\pm 0.07$ & \ldots& \ldots \\
\ \     +Planck+WP+BAO            & $3.5\pm0.3$    & $1.66\pm0.06$ & $0.73\pm0.12$& $0.20\pm 0.07$ & \ldots& \ldots \\
\citet{dietrich19}                & $5.58^{+0.96}_{-1.46}$ & $1.650^{+0.097}_{-0.096}$ & $1.27^{+0.47}_{-0.51}$ &  $0.173^{+0.073}_{-0.052}$ & \ldots & \ldots \\
\multicolumn{3}{l}{\textsc{\bf DES-SV WL Shear}} & & & &\\
\ \ priors & \ldots               & $1.67\pm0.08$ & $0.55\pm0.32$ &$0.20\pm0.07$ & $0.94\pm 0.18$ & $0.25\pm 0.12$\\
\ \ \ \ with $B_\mathrm{SZ}$ prior & $12.0_{-6.7}^{+2.6}$ & $1.65_{-0.09}^{+0.08}$ & $0.50_{-0.30}^{+0.31}$ & $0.20\pm0.07$ &$0.94^{+0.17}_{-0.18}$ & $0.24^{+0.11}_{-0.12}$\\
\ \ \ \ free $B_\mathrm{SZ}$       & $10.8_{-5.2}^{+2.3}$ & $1.30_{-0.44}^{+0.22}$ & $0.50\pm0.32$       &$0.20\pm0.07$  &$0.94\pm0.18$       &$0.24_{-0.12}^{+0.10}$\\
\hline\hline
 \end{tabular}
\label{tab:ScalRel}
\end{table*}

%% file: affiliations.tex
$^{1}$ Faculty of Physics, Ludwig-Maximilians-Universit\"at, Scheinerstr. 1, 81679 Munich, Germany\\
$^{2}$ Excellence Cluster Universe, Boltzmannstr.\ 2, 85748 Garching, Germany\\
$^{3}$ Argonne National Laboratory, 9700 South Cass Avenue, Lemont, IL 60439, USA\\
$^{4}$ Kavli Institute for Cosmological Physics, University of Chicago, 5640 South Ellis Avenue, Chicago, IL 60637\\
$^{5}$ Argelander-Institut f\"ur Astronomie, Universit\"at Bonn, Auf dem H\"ugel 71, 53121, Bonn, Germany\\
$^{6}$ Max Planck Institute for Extraterrestrial Physics, Giessenbachstrasse, 85748 Garching, Germany\\
$^{7}$ Jodrell Bank Center for Astrophysics, School of Physics and Astronomy, University of Manchester, Oxford Road, Manchester, M13 9PL, UK\\
$^{8}$ Department of Astronomy, University of Illinois at Urbana-Champaign, 1002 W. Green Street, Urbana, IL 61801, USA\\
$^{9}$ National Center for Supercomputing Applications, 1205 West Clark St., Urbana, IL 61801, USA\\
$^{10}$ Kavli Institute for Particle Astrophysics \& Cosmology, P. O. Box 2450, Stanford University, Stanford, CA 94305, USA\\
$^{11}$ SLAC National Accelerator Laboratory, Menlo Park, CA 94025, USA\\
$^{12}$ Department of Physics and Astronomy, University of Pennsylvania, Philadelphia, PA 19104, USA\\
$^{13}$ Department of Physics, ETH Zurich, Wolfgang-Pauli-Strasse 16, CH-8093 Zurich, Switzerland\\
$^{14}$ INAF-Osservatorio Astronomico di Trieste, via G. B. Tiepolo 11, I-34143 Trieste, Italy\\
$^{15}$ Brookhaven National Laboratory, Bldg 510, Upton, NY 11973, USA\\
$^{16}$ Center for Cosmology and Astro-Particle Physics, The Ohio State University, Columbus, OH 43210, USA\\
$^{17}$ Department of Physics, The Ohio State University, Columbus, OH 43210, USA\\
$^{18}$ Institute for Astronomy, University of Edinburgh, Edinburgh EH9 3HJ, UK\\
$^{19}$ Fermi National Accelerator Laboratory, P. O. Box 500, Batavia, IL 60510, USA\\
$^{20}$ Department of Astronomy and Astrophysics, University of Chicago, 5640 South Ellis Avenue, Chicago, IL 60637\\
$^{21}$ Kavli Institute for Cosmological Physics, University of Chicago, Chicago, IL 60637, USA\\
$^{22}$ Department of Physics, IIT Hyderabad, Kandi, Telangana 502285, India\\
$^{23}$ Center for Astrophysics and Space Astronomy, Department of Astrophysical and Planetary Science, University of Colorado, Boulder, C0 80309, USA\\
$^{24}$ NASA Ames Research Center, Moffett Field, CA 94035, USA\\
$^{25}$ School of Physics, University of Melbourne, Parkville, VIC 3010, Australia\\
$^{26}$ Physics Department, Case Western Reserve University, Cleveland, Ohio 44106, USA\\
$^{27}$ Cerro Tololo Inter-American Observatory, National Optical Astronomy Observatory, Casilla 603, La Serena, Chile\\
$^{28}$ Department of Physics \& Astronomy, University College London, Gower Street, London, WC1E 6BT, UK\\
$^{29}$ Department of Physics and Electronics, Rhodes University, PO Box 94, Grahamstown, 6140, South Africa\\
$^{30}$ Institute of Cosmology \& Gravitation, University of Portsmouth, Portsmouth, PO1 3FX, UK\\
$^{31}$ CNRS, UMR 7095, Institut d'Astrophysique de Paris, F-75014, Paris, France\\
$^{32}$ Sorbonne Universit\'es, UPMC Univ Paris 06, UMR 7095, Institut d'Astrophysique de Paris, F-75014, Paris, France\\
$^{33}$ Laborat\'orio Interinstitucional de e-Astronomia - LIneA, Rua Gal. Jos\'e Cristino 77, Rio de Janeiro, RJ - 20921-400, Brazil\\
$^{34}$ Observat\'orio Nacional, Rua Gal. Jos\'e Cristino 77, Rio de Janeiro, RJ - 20921-400, Brazil\\
$^{35}$ Institut de F\'{\i}sica d'Altes Energies (IFAE), The Barcelona Institute of Science and Technology, Campus UAB, 08193 Bellaterra (Barcelona) Spain\\
$^{36}$ Institut d'Estudis Espacials de Catalunya (IEEC), 08193 Barcelona, Spain\\
$^{37}$ Institute of Space Sciences (ICE, CSIC),  Campus UAB, Carrer de Can Magrans, s/n,  08193 Barcelona, Spain\\
$^{38}$ Centro de Investigaciones Energ\'eticas, Medioambientales y Tecnol\'ogicas (CIEMAT), Madrid, Spain\\
$^{39}$ Department of Astronomy, University of Michigan, Ann Arbor, MI 48109, USA\\
$^{40}$ Department of Physics, University of Michigan, Ann Arbor, MI 48109, USA\\
$^{41}$ Instituto de Fisica Teorica UAM/CSIC, Universidad Autonoma de Madrid, 28049 Madrid, Spain\\
$^{42}$ Santa Cruz Institute for Particle Physics, Santa Cruz, CA 95064, USA\\
$^{43}$ Australian Astronomical Observatory, North Ryde, NSW 2113, Australia\\
$^{44}$ Departamento de F\'isica Matem\'atica, Instituto de F\'isica, Universidade de S\~ao Paulo, CP 66318, S\~ao Paulo, SP, 05314-970, Brazil\\
$^{45}$ Department of Astrophysical Sciences, Princeton University, Peyton Hall, Princeton, NJ 08544, USA\\
$^{46}$ Instituci\'o Catalana de Recerca i Estudis Avan\c{c}ats, E-08010 Barcelona, Spain\\
$^{47}$ Jet Propulsion Laboratory, California Institute of Technology, 4800 Oak Grove Dr., Pasadena, CA 91109, USA\\
$^{48}$ Department of Physics and Astronomy, Pevensey Building, University of Sussex, Brighton, BN1 9QH, UK\\
$^{49}$ School of Physics and Astronomy, University of Southampton,  Southampton, SO17 1BJ, UK\\
$^{50}$ Instituto de F\'isica Gleb Wataghin, Universidade Estadual de Campinas, 13083-859, Campinas, SP, Brazil\\
$^{51}$ Computer Science and Mathematics Division, Oak Ridge National Laboratory, Oak Ridge, TN 37831\\